\documentclass[useAMS,usenatbib]{mnras}
\usepackage{graphicx}
\usepackage{dblfloatfix}
\usepackage{amsmath}
\usepackage{amssymb}
\usepackage{natbib}
\usepackage[usenames, dvipsnames]{xcolor}
\usepackage{pdflscape}
\usepackage{longtable}
\usepackage[bottom]{footmisc}
\usepackage[flushleft]{threeparttable}

\newcommand{\bassic}{HB13}
\newcommand{\ellisa}{{\textsc ELLISA}}

\defcitealias{Mateu}{M12}

\title[QUEST Eclipsing Binaries and ELLISA]{Eclipsing binary search in the QUEST low latitude catalogue and the ELLISA light curve simulator}
\author[B. Cuevas-Otahola et al.]
  {Bolivia~Cuevas-Otahola$^{1,}$$^{2,}$$^3$, Cecilia~Mateu$^{1,4}$, Fabiola~Hern\'andez-P\'erez$^2$, \newauthor
  Juan~Jos\'e~Downes$^{1,5}$, A. Katherina~Vivas$^6$ and
  C\'esar~Brice\~no$^6$\\
  $^1$Centro de Investigaciones de Astronom\'ia, AP 264, M\'erida 5101-A, Venezuela\\
  $^2$Universidad de Los Andes, Facultad de Ciencias, Departamento de F\'isica, M\'erida, Venezuela\\
  $^3$Instituto Nacional de Astrof\'isica, \'Optica y Electr\'onica, Ap. 51, 72000 Puebla, M\'exico, M\'exico\\
  $^4$Departamento de Astronom\'{i}a, Instituto de F\'{i}sica, Universidad de la Rep\'{u}blica, Igu\'a 4225, 11400 Montevideo, Uruguay \\
  $^5$PDU Ciencias F\'{i}sicas, Centro Universitario Regional del Este (CURE), Universidad de la Rep\'{u}blica, 27000 Rocha, Uruguay \\ 
  $^6$Cerro Tololo Inter-American Observatory, National Optical Astronomy Observatory, Casilla 603, La Serena, 1700000, Chile}
\date{Released 2019 Xxxxx XX}

\def\LaTeX{L\kern-.36em\raise.3ex\hbox{a}\kern-.15em
    T\kern-.1667em\lower.7ex\hbox{E}\kern-.125emX}

\begin{document}

\label{firstpage}

\maketitle

\begin{abstract}
The realistic simulation of variable star populations is fundamental
to determine the selection function and contamination in existing and
upcoming multi-epoch surveys.
We present \ellisa, a simulator that produces an ensemble of mock light
curves for a population of eclipsing binaries obtained from
physical and orbital parameters consistent with different galactic
populations, and which considers user-supplied time sampling and photometric
errors to represent any given survey.
We carried out a search for eclipsing binaries in the QUEST low-galactic latitude
catalogue of variable stars, spanning an area of 476 sq. deg at
$-25^\circ \la b \la 30^\circ$ and
$190^ \circ \leqslant l \leqslant 230^ \circ$ towards the galactic
anti-centre, and use \ellisa~to characterise the completeness of the
resulting catalogue in terms of amplitudes and periods of variation
as well as eclipsing binary type.
The resulting catalogue consists of $1,125$ eclipsing binaries,
out of which $179$, $60$ and $886$ are EA, EB and
EW types respectively. We estimate, on average, $30\%$
completeness in the period range $0.25 \la P/d \la 1$ for
EB+EW binaries and $15\%$ completeness for EA
binaries with periods $2 \la P/d \la 10$, being the time
sampling the primary factor determining the completeness
of each type of eclipsing binary.
This is one of few eclipsing binary catalogues reported with an estimate of the selection function. 
Mock eclipsing binary light curve libraries produced with
\ellisa~can be used to estimate the selection function and optimise  
eclipsing binary searches in upcoming multi-epoch surveys such as 
 Gaia, the Panoramic Survey Telescope and Rapid Response System, 
the Zwicky Transient Factory or the Large Synoptic Survey Telescope.
\end{abstract}

\begin{keywords}
(stars:) binaries: eclipsing, astronomical data bases: catalogues, methods: numerical
\end{keywords}

\section{Introduction}

Eclipsing binaries are widely recognised for their importance in different astrophysical areas. In variability surveys, they are usually the most abundant type of variable star, particularly at low Galactic latitude. Although the specific proportions depend on each survey's depth, cadence and area probed (particularly galactic latitude), in the All Sky Automated Survey \citep[ASAS,][]{Pojmanski} and Catalina Real-Time Transient Survey \citep[CRTS,]{Drake} they amount to 25\% and 58\% percent of all variable stars respectively.

Eclipsing contact binaries or W-UMa type variables (EWs), in particular, follow a Period-Luminosity-Colour relation \citep{Rucinskipcolor,Rucinski1997}, which implies they can serve as standard candles for distance measurements \citep{Rucinski1996}. The latest estimates give an error of $\pm0.25$ mag in distance modulus for these stars, which corresponds to a distance uncertainty of $\sim12\%$ \citep{Rucinski2004}. 
Although significantly fainter ($M_V>2$) than other, more traditional, standard candles such as RR Lyraes ($M_V\sim+0.6$) and Cepheids ($M_V<-2$), EWs are numerous and ubiquitous as they trace populations of any age and metallicity. 
The advent of deep all-sky multi-epoch surveys such as Gaia, 
 the Panoramic Survey Telescope and Rapid Response System (PanSTARRS), 
the Zwicky Transient Factory (ZTF) or the Large Synoptic Survey Telescope (LSST) \citep{Gaia2017b,Kaiser2010,Ivezic2010,Smith2014}, together with the first 3D extinction maps \citep{Sale2014,Green2015}, will open up the opportunity to use EWs as tracers of Galactic structure for the first time.  For example, Gaia will be capable of observing an $M_V=3$ EW star up to $\sim30$ kpc without extinction, or up to $\sim15$ kpc with $A_V=1$, effectively probing a considerable volume of the Galactic Disc and inner Halo.  

Numerous works have shown many potential uses for standard candles as probes of Galactic structure, e.g. to discover and trace stellar overdensities, clouds and tidal streams \citep[e.g.][]{Vivas2006,Sesar2010p,Baker2015}, trace radial and vertical metallicity gradients \citep{Luck2006} and to measure the density profiles of different Galactic components \citep[e.g.][]{Brown2008,Vivas2006,Sesar,Cohen2017}. However, for any tracer catalogue to be useful for studies of the Galactic density profiles, a thorough understanding of its completeness is required, which needs to be estimated through simulations. In RR Lyrae surveys, light curve simulations have been done extensively to model the survey completeness.
 \citet{Vivas2004,Miceli2008}; \citet[][hereafter \citetalias{Mateu}]{Mateu}; \citet{Sesar2017b}, for example, produce synthetic light curve catalogues for a population of mock RR Lyrae stars, which are then run through the period-finding and RR Lyrae identification algorithms used to characterise the survey's completeness by looking at the fraction of recovered stars as a function of different parameters, such as magnitude, amplitude and number of observations. This completeness in the identification of RR Lyrae stars can range from $\sim60\%$ in the SEKBO and Catalina surveys \citep{Keller2008,Drake2013,Torrealba2015} to as high as $>90\%$ in LONEOS, SDSS Stripe 82, PS1 or QUEST \citep[][\citetalias{Mateu}]{Sesar2010p,Sesar2017b,Vivas2004}. Estimates such as these, however, have not been provided to date for eclipsing binary surveys.

The fact that eclipsing binaries are the most common type of variable star also means they are a very common contaminant in surveys for other types of variables. EWs are frequent contaminants of RRc and Delta Scuti surveys \citep[e.g.][\citetalias{Mateu}]{Vivas2004,Kinman2010}. Their light curve period ranges are similar (from a few hours to $\lesssim1.5$~d), and when time sampling is irregular and observations sparse, the light curve shapes can be difficult to tell apart and period aliasing can be a source of confusion.  

Although currently many codes exist to simulate eclipsing binary light curves, such as NightFall \citep{Wichmann2011}, Wilson-Devinney \citep[WD,][]{Wilson} and EBOP \citep[Eclipsing Binary Orbit Program,][]{Etzel}, these are oriented towards simulating light curves for individual binary systems in great detail. However, as a population, eclipsing binary light curves are difficult to model because stars in almost any two evolutionary stages can be part of a binary system and the proportions of the different types of stars paired have to be modelled consistently with the initial mass function and star formation history of the population, and the effects of mass transfer on the stellar evolution of each component \citep[see e.g.][]{Hernandez}.  \citet{Prsa2011} approached this problem to estimate the eclipsing binary yield of LSST by simulating a library of eclipsing binary light curves with physical and orbital parameters drawn at random from a set of given distributions, which were then fed into PHOEBE \citep{Prsa2005}, a code based on WD. 

Our goal in this work was, therefore, to develop a tool to simulate light curves for populations of eclipsing binaries. \ellisa~(Eclipsing binary Light curve LIbrary SimulAtor), is based on stellar population synthesis models \citep{Hernandez} to produce binary systems with physical characteristics and in numbers consistent with the parent stellar population, and reproducing the observational characteristics of a given survey: time sampling, typical photometric errors in each filter, bright and faint magnitude limits, and so on (Section~\ref{Sec:ellisa}). 
The light curve catalogues simulated with \ellisa~ will allow characterizing the completeness and possible biases of any eclipsing binary search, serve as benchmarks for the optimization of eclipsing binary observing strategies and search algorithms, and provide estimates of the expected levels of contamination of searches for other types of variable stars. Finally, we use the QUEST low latitude catalogue of variable stars \citepalias{Mateu} as a case study, and implement \ellisa~to guide the search and characterise the completeness of the resulting eclipsing binary catalogue. The \ellisa~code is publicly available at a GitHub repository\footnote{\url{https://github.com/umbramortem/ELLISA}}.

This paper is organized as follows: Section \ref{Sec:ellisa} describes the \ellisa~code, which is used to generate a mock catalogue to guide the search for eclipsing binaries in the QUEST catalogue of variable stars described in Section \ref{binary_search}.  Using the ELLISA mock catalogue, the completeness of the catalogue obtained is characterised in Section \ref{s:compsec} as a function of amplitude, magnitude and spatial distribution, for each eclipsing binary type.  Section \ref{sum_and_con} contains summary and conclusions.
 
\section{\ellisa: Eclipsing binary Light curve LIbrary SimulAtor}\label{Sec:ellisa}

The goal of the \ellisa~simulator is to generate a library of synthetic multi-filter light curves for a population of eclipsing binaries, reproducing the time sampling and photometric errors representative of the survey the user wants to simulate.  This makes it possible to generate a synthetic library that mimics the way we would observe a population of eclipsing binaries with an arbitrary instrument and time sampling. \ellisa~is made publicly available as a  {\texttt Python} stand-alone code and library at a GitHub repository\footnotemark[\value{footnote}].

\begin{figure*}
\begin{center}
\includegraphics[width=\textwidth]{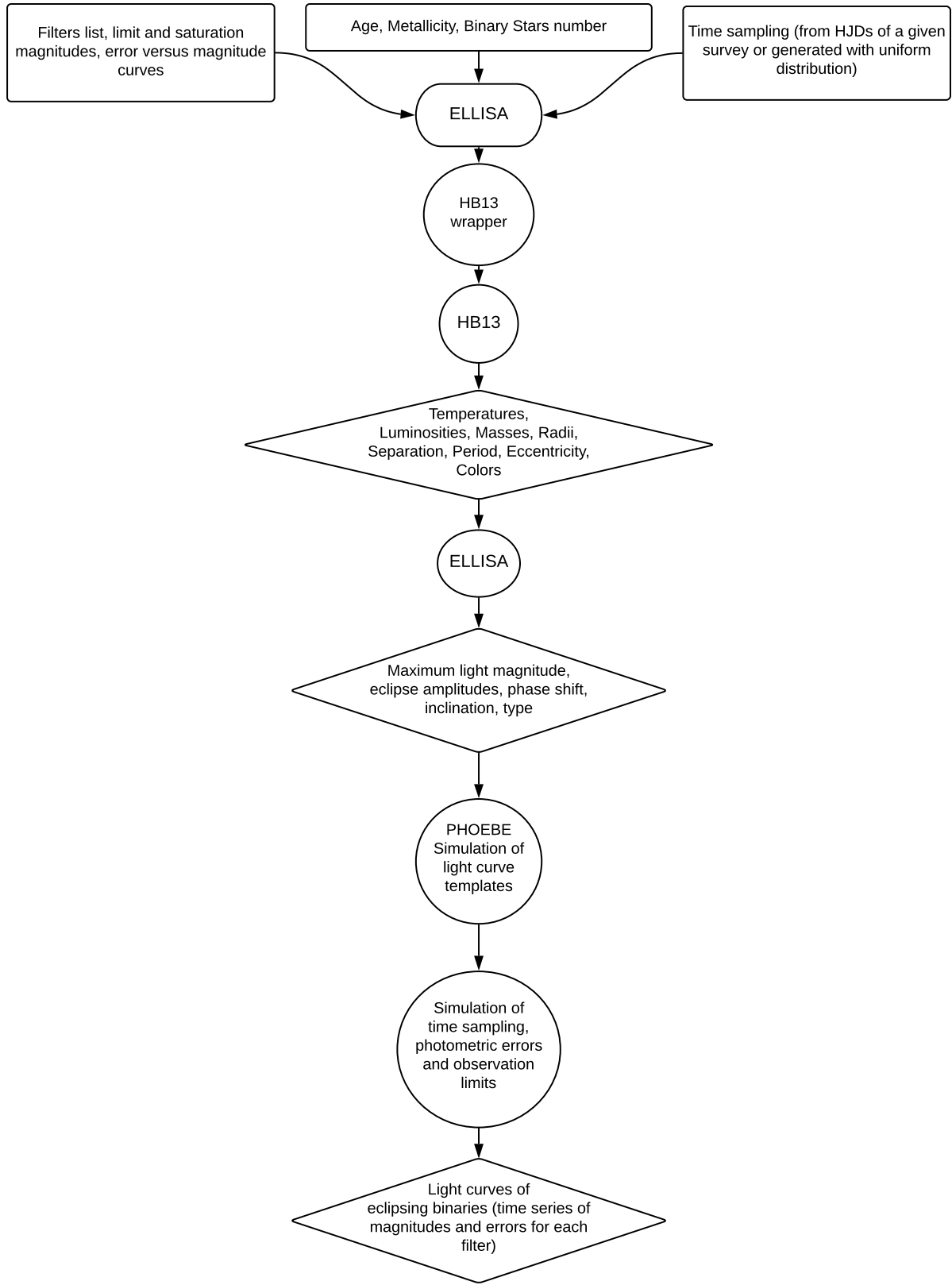}
\caption{\ellisa~flowchart. Rectangles, ellipses/circles and diamonds indicate user-supplied input, subroutines and outputs respectively.}
\label{flujogramasimulador}
\end{center}
\end{figure*}

The code's structure is shown in Figure \ref{flujogramasimulador}, where user inputs are shown with squares, 
subroutines used by \ellisa~are shown in circles and the outputs produced are shown with rhombuses.  The overall process is as follows. The user supplies \ellisa~the stellar population type, filter set and number of binary systems to simulate.  This data is used by a routine that works as a wrapper for the binary stellar population synthesis code  HB13 by \cite{Hernandez}, which simulates a synthetic population of binary systems at zero-age of a simple stellar population (SSP), calculates its stellar evolution according to the selected initial mass function,
and returns the physical parameters, magnitudes and colours of the individual stars in the binary population. 

From these parameters the type of eclipsing binary (EW,EB and EA) is identified and, for each filter, light curve parameters --eclipse amplitudes, colours, maximum light magnitudes-- are computed using the widely known code PHOEBE \citep{Phoebe}.  Then the observation process is simulated by sampling the light curve in the specific epochs given in the input time sampling, and the photometric errors and observation limits are simulated based on the (user-supplied) survey characteristics.  Finally, the simulator returns the time series containing the magnitudes, photometric errors and observation dates for each binary system, in each filter selected by the user. 
In what follows we describe in detail each of these steps.

\subsection{Simulation of Physical Parameters of Binaries using \bassic}\label{Sec:parfisicosbassic}

In this part, \ellisa~uses the \bassic~code from \citep{Hernandez} to generate the physical and orbital parameters --masses, separations, effective temperature, etc. --  of the binary population, consistently with the stellar evolution of the population being simulated.

\bassic~is a code developed by \citet{Hernandez} that generates and follows the stellar evolution of an SSP, consisting of isolated stars and binary systems.  This is implemented as a \texttt{Fortran} code, bundled and run inside \ellisa~through a \texttt{Python} wrapper subroutine. Here we summarise the procedures followed by \bassic, and we refer the reader to \citet{Hernandez} for further details.

\paragraph*{Stellar population selection}

The wrapper subroutine allows simulating predefined simple and composite stellar populations.  In its current version, \ellisa~allows the user to choose one of the following options, tailored to resemble the main Galactic components: Halo type, Bulge type, Thick disc type, Thin disc type.

The ages and metallicities used for the preset populations were chosen following \citep[][their Table 1]{Robin}. The Halo, Bulge and Thick Disc populations are simulated using SSPs, and the extended star formation history of the Thin Disc is simulated using 7 SSPs with ages between 100 Myr to 10 Gyr and metallicities between -0.12 dex and 0.01 dex, given by \citet{Robin}.  These SSPs are combined in equal proportions (by number) for each age bin, so as to approximately reproduce the star formation history of the Thin Disc with the local density reported by \cite{Robin} for each age bin.
 
\paragraph*{Initial Physical and Orbital Parameters}

The physical and orbital parameters are simulated for the binary population at zero-age, as follows:

\begin{itemize}

\item Stellar masses ($M_1$) are randomly drawn from a \citet{Chabrier} Initial Mass Function ranging from $0.1M_\odot$ to $100M_\odot$, parametrized as a log-normal distribution with characteristic mass $\log m_c=0.08$ and variance $\sigma^2=<(\log m - <\log m>)^2>=0.47$. 

\item For each potential primary star, a binary probability is drawn at random as a function of the spectral type, based on the compilation by \citet{Lada2006} summarized in  Table \ref{t:lada}. This reproduces the larger binary probability observed for more massive stars. It is important to stress that these correspond to binary fractions at zero age, which change as the stellar population evolves and the stellar evolution under the effects of mass transfer is taken into account. For example, the fraction of F-G primary stars in binaries changes from 13.7\% at zero-age to 8.6\% at 10~Gyr.

\item For the stars randomly assigned to binary systems in the previous step, secondary masses are computed from a mass ratio $q$ drawn at random from a uniform distribution ranging from 0 to 1.  \cite{Reggiani2011} and \cite{Reggiani2013} find that distribution of mass ratios is consistent with being uniform $(dN/dq=q^{0.25\pm0.29})$ in the Galactic field, as well as in many star forming regions. They also find the mass ratio distribution to be independent of the binary separation.

\item Orbital periods are drawn at random from the \citet{Duquennoy} log-normal distribution with mean log-period  $\overline{\log P(d)} = 4.4$ and standard deviation $\sigma_{\log P} =2.3$. Orbital semi-major axes are computed from Kepler's Third Law $GMa^{-3}= 4\pi^2P^{-2}$.

\item Initial orbital eccentricities are randomly drawn from a uniform distribution ranging from 0 to 1.

\end{itemize}

\paragraph*{Binary Evolution}

Having generated the orbital and physical parameters, the next step is to follow the stellar evolution of each binary system. \bassic~uses the \citet{Hurley2002} evolutionary tracks in order to follow the evolution of each binary, taking into account the degree of mass transfer in each evolutionary stage, based on the stellar masses and radii, and  the orbital parameters of the system.  The evolutionary stages followed go from the zero-age main sequence to the remnant stages (black hole, neutron star or white dwarf) for stars with initial masses ranging from 0.1 to a 100 $M_\odot$ and metallicities from Z=0.0004 to Z=0.01.  In addition, to cover evolutionary aspects of He white dwarf stars with collisions considered by \citet{Hurley2002} as destroyed systems,  \citet{Hernandez} use the model proposed by \citet{Han2002} which assumes Extreme Horizontal Branch stars (EHB) are formed through this channel.  The isochrones used in the model are built from the evolutionary tracks described in \citet{Hernandez}.

The \citeauthor{Hurley2002} evolutionary tracks code allows computing the evolution of a binary system at any age, returning the temperature and luminosity for each star in the binary. Absolute magnitudes are computed using the BaSeL 3.1 \citep{Westera2002} spectral library.  Currently the photometric systems available are Johnson-Cousins (UBVRIJHK), SDSS (\emph{ugriz}) and HST (F814W, F775W, F625W, F606W, F555W, F445W, F435W, F410W, F330W, F250W, F220W).

\begin{table}%[htdp]
\caption{Binary fraction as function of spectral type}
\begin{center}
\begin{tabular}{ccc}
\hline
Spectral type & Binary fraction  & Reference\\
\hline
O	&	0.72	& Mason et al. 1998\\
O-B	&	0.65	& Preibisch et al. 1999\\
B-A	&	0.62 $\pm$ 0.2	& Patience et al. 2002\\
G-K	&	0.58$\pm$0.1	&	Duquennoy \& Mayor 1991\\
M	&	0.49$\pm$0.09	&	Fischer \& Marcy 1992\\
Late-type M	&	0.26$\pm$0.1	& Basri \& Reiners 2006\\
\hline
\label{t:lada}
\end{tabular}
\end{center}
\end{table}

\subsection{Calculation of light curve parameters using PHOEBE}\label{s:phoe_sec}

Having generated the physical and orbital parameters for the binaries, \ellisa~goes on to determine the light curve parameters (primary and secondary eclipse amplitudes, maximum light magnitude) corresponding to each system of the simulated populations.  The first step is to determine the light curve template to be used.

\paragraph*{Light curves}

To generate the light curves we use PHOEBE (PHysics Of Eclipsing BinariEs) \citep{Phoebe}, a modelling package for eclipsing binaries based on the well-known Wilson-Devinney code \citep[WD,][]{Wilson}.  We used the Legacy version of PHOEBE\footnote{\url{http://phoebe-project.org/1.0}}, with {\sc Python} implementation capabilities.  We make use of the LC PHOEBE feature which employs the LC WD's program to generate the light curve based on the binary's physical and orbital parameters. 

For this work we use the PHOEBE functionality for contact and detached systems. Therefore, two types of models are used to simulate the shapes of the eclipsing binary light curves: one for detached or Algol type, which hereinafter will be referred as EA systems, and another for semi-detached ($\beta$Lyrae) or contact (WUMa) systems, which in what follows will be referred as EB+EW systems \citep[following the notation of the GCVS - General Catalog of Variable Stars][]{Samus}. The first step is to figure out whether the system is completely detached or not. For this, we follow \citet{Eggletonbook} and for each binary we compute the critical period $P_{\rm crit}$, i.e. the shortest period that a system with a certain mass ratio should have to make mass transfer events possible without overflowing its Roche lobe.  According to \citet{Eggletonbook} (their Eq. 3.10) the critical period is given by

\begin{equation}
P_{\rm crit} \sim 0.35 \sqrt{\frac{R^3}{M_1}} \left( \frac{2}{1+q} \right) ^{0.2} ,
\end{equation}

\noindent where $M_1$ and $R$ denote the mass and radius of the primary and $q$ the system's mass ratio.\\

Through this criterion, we will separate the systems generated by \bassic~into two blocks.   The first group is constituted by EA systems whose separation is large enough to prevent mass transfer.  A binary is of EA type if it has a period $P<P_{\rm crit}$. The second, and complementary group, contains those systems with one or the two stars transferring mass to its companion, which correspond in this case to EB or EW light curves. 

In order to shape the light curve on its entirety, given its type, the following orbital parameters are passed to PHOEBE to perform the orbit calculation:

\begin{itemize}
\item{Initial inclination angle $i$ of the orbital plane for each system, randomly drawn from a $\cos (i)$, $i \in [0,\pi]$ distribution \citep[see e.g.][]{Arenou}. The range of inclination angles in which a given binary will be detected as eclipsing is given by equation \ref{incleq}.  This holds for EA, EW and EB systems.  For shorter semi-major axes the range of inclination values is larger as we can see in Figure} \ref{inclinacionfig}

\begin{equation}
a \cos (i) < R_1+R_2.
\label{incleq}
\end{equation}

\begin{figure}
\begin{center}
\includegraphics[width=0.7\columnwidth]{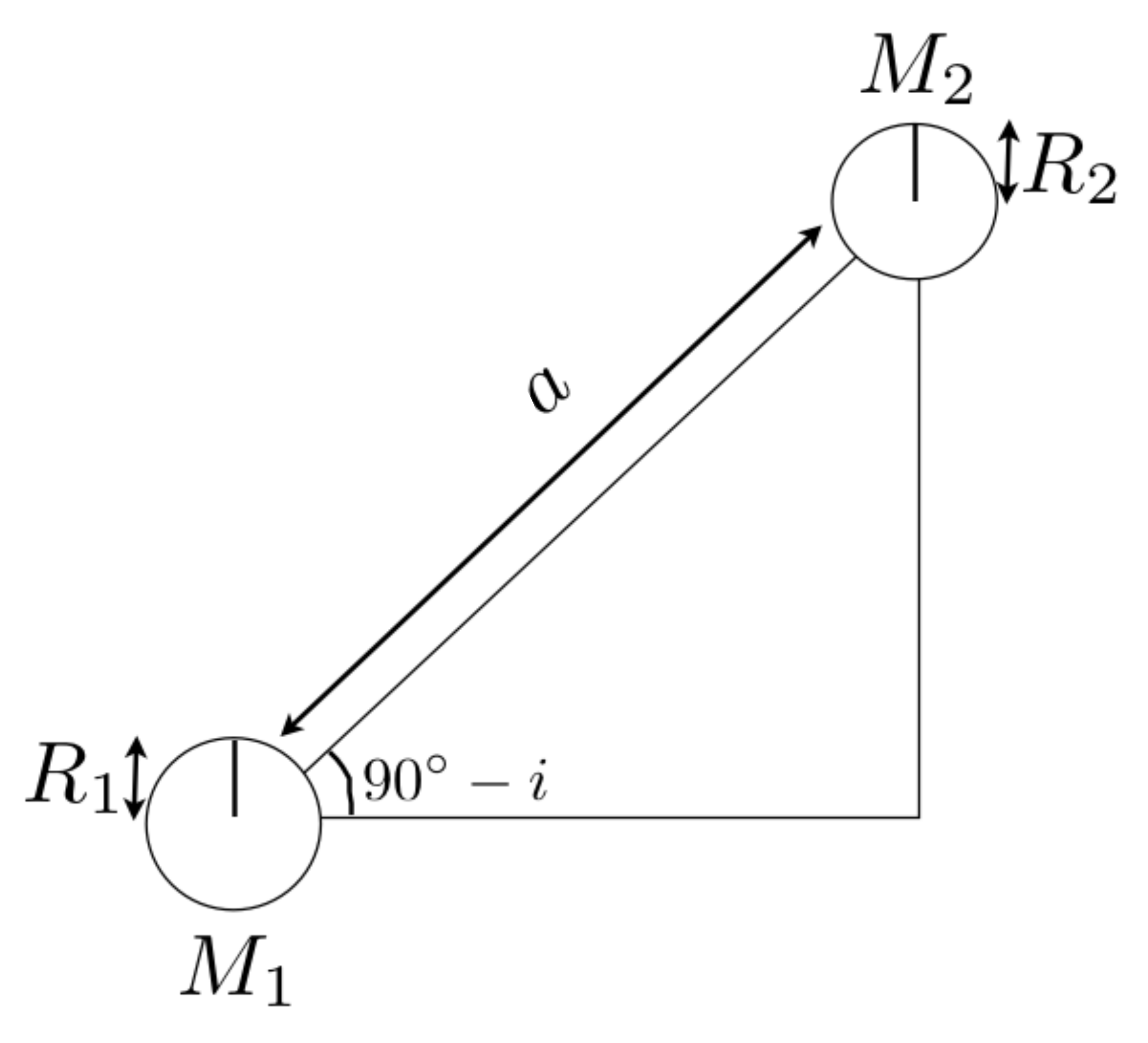}
\caption{Schematic representation of an eclipsing binary.  Inclination-radius relation of the components.}
\label{inclinacionfig}
\end{center}
\end{figure}

\item{Eccentricity, period and semi-major axis distributions, given by HB13.}

\item{Synchronicity parameter which is the ratio between rotation and orbital angular velocity, follows equation \ref{sync} when differential rotation is neglected
\begin{equation}
F=\sqrt{\frac{1+\epsilon}{(1-\epsilon)^3}}.
\label{sync}
\end{equation}}

\noindent where $\epsilon$ is the orbital eccentricity.

\item{Argument of the periastron, randomly drawn from a uniform distribution in the range $[0,2\pi]$}
\end{itemize}

Once the orbit is added to the simulated system, the stellar parameters are also passed to PHOEBE in order to obtain geometrical light curves parameters.

Limb darkening effects are included in every system computation and are given as function of the selected passband in the case of the WD feature or can be dynamically computed from predefined tables (See PHOEBE scientific reference for further details).

The light curves physical quantities are given in units of flux, computed from the given passband luminosities in the available passbands systems (Stromgren, Johnson, Gaia, Tycho, Kepler, Hipparcos, LSST, Cousins).

\paragraph*{Apparent magnitudes}
At this point we transform the absolute magnitudes to apparent ones by adding a distance modulus. For this, the user selects a reference band in which the final distribution of mean apparent magnitudes will be uniform and limited at the bright and faint ends by the user-supplied saturation and limiting magnitudes, respectively. The apparent magnitude in the reference band is drawn at random from this uniform distribution and the corresponding distance modules is computed from the simulated absolute magnitude. This distance modulus is used to transform the  magnitudes in the remaining bands from absolute to apparent. Finally, observations outside the (user-supplied) saturation and limiting magnitudes of each band are discarded.

\subsection{Time sampling and photometric errors}

The next step is to simulate the observation process, reproducing the time sampling, photometric errors and bright and faint limits of the survey for all the filters.

\paragraph*{Time sampling}

The time sampling is generated by \ellisa~based on user-supplied files containing the heliocentric julian dates (HJDs) of the observations in each filter.

The (error-free) magnitude $m_i^{\rm F0}$ is obtained by evaluating the light curve template for each filter $\rm F$ at the phase $\phi_i^{\rm F}$, corresponding to the $t_i^{\rm F}$ observation epoch, given by the following expression 

\begin{equation}
\phi_i^{\rm F}=\frac{t_i^{\rm F}}{P} - int \bigg{(}\frac{t_i^{\rm F}}{P}\bigg{)} + \phi_{\rm off}  
\end{equation}  

\noindent where $t_i^{\rm F}$ is the HJD of the $i$-th observation in the $\rm F$ filter, $P$ is the light curve period and $\phi_{\rm off}$ is the phase shift.  The phase shift $\phi_{\rm off}$ is randomly drawn from a uniform distribution in the range $\phi_{\rm off}\in[0,1)$. The  phase-offset is filter-independent and it is taken as the phase in which the primary eclipse occurs.

\paragraph*{Photometric errors}

\begin{figure}
\begin{center}
\includegraphics[width=\columnwidth]{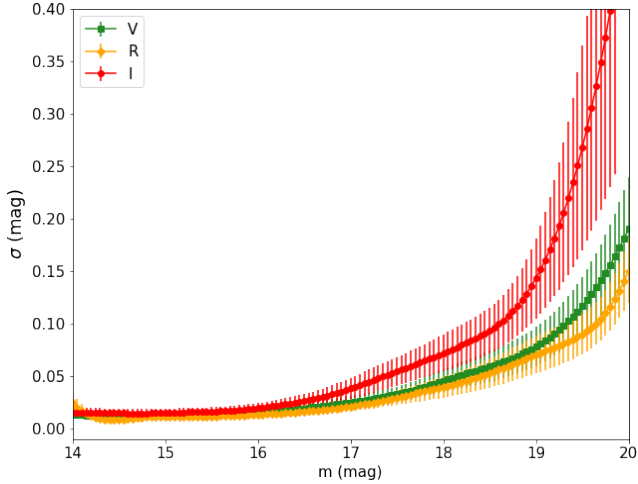}
\caption{Error versus magnitude curves for QUEST catalogue in the VRI Johnson-Cousins filters, the bars represent the mean standard deviation $\delta$ of the typical photometric error $\sigma$ at each magnitude.}
\label{errvsmag}
\end{center}
\end{figure}

\begin{figure*}
\begin{center}
\includegraphics[width=2\columnwidth]{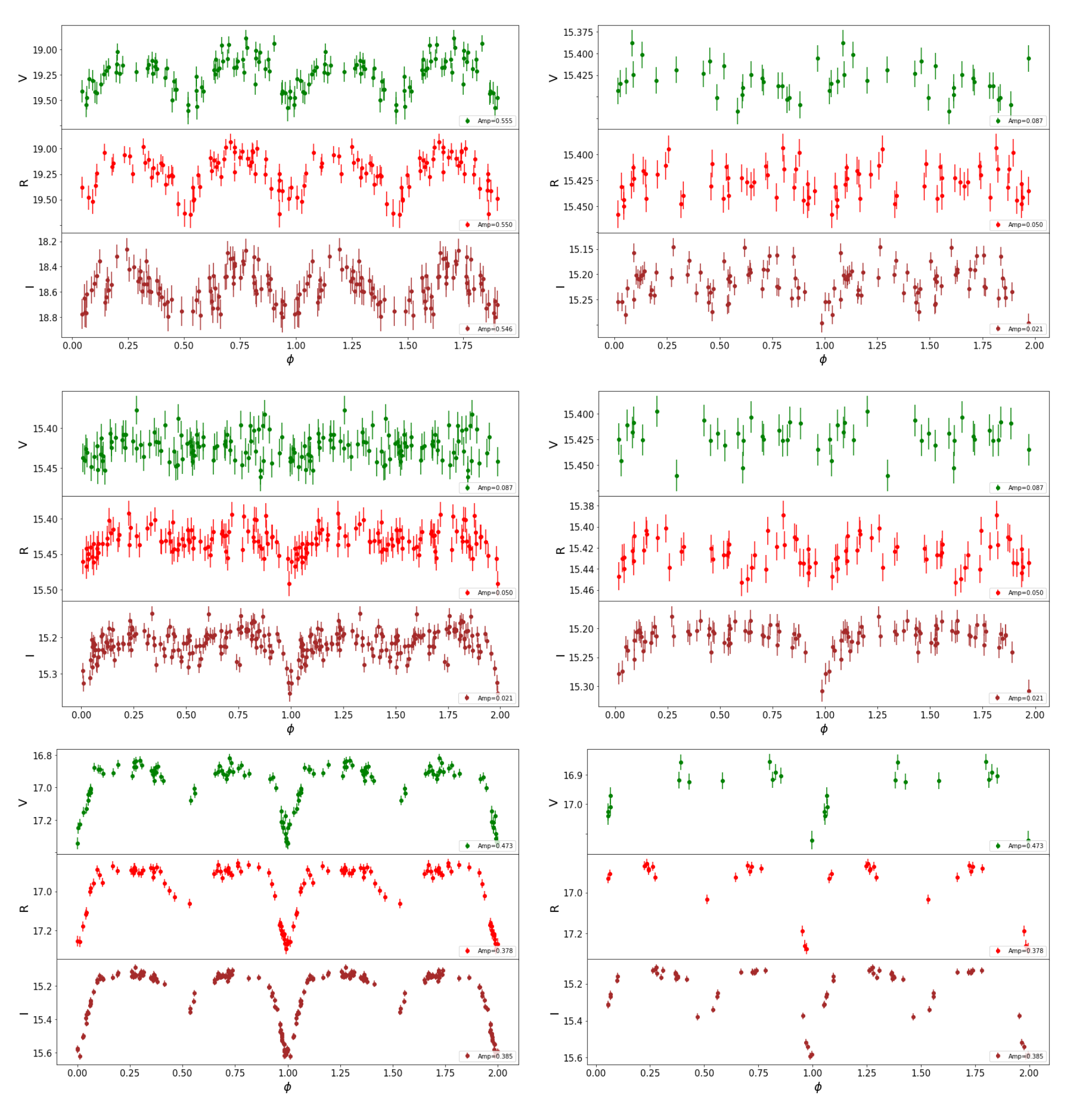}
\caption{Simulated QUEST VRI light curves:  EW system (upper panels), EB system (middle panels) and EA system (bottom panels) with very dense (left panels) and very sparse (right panels) time sampling .}
\label{curves}
\end{center}
\end{figure*}

The photometric errors are simulated as a function of magnitude, based on user-supplied error versus magnitude curves characteristic of the survey that is being simulated. As an example, Figure \ref{errvsmag} shows error versus magnitude curves for the QUEST survey in the VRI filters. These curves represent the mean standard deviation $\sigma$ (in magnitudes) of non-variable stars per magnitude bin \citep[][\citetalias{Mateu}]{Vivas2004}. The error bars are indicative of the actual dispersion of multiple measurements for non-variable stars at each magnitude bin, which we call $\delta$, and represents the dispersion of the photometric error $\sigma$ at that magnitude. The curves in Figure \ref{errvsmag} were computed for the QUEST low-latitude catalogue of variable stars \citepalias{Mateu}. 

The error-convolved magnitude is thus computed as $m_i=m_i^{\rm F0} + \Delta m_i^{\rm F}$, where $\Delta m_i^{\rm F}$ is a random number drawn from a gaussian distribution $G(0,\sigma)$ with zero-mean and standard deviation $\sigma$, which is, in turn, drawn from a gaussian $G(0,\delta)$ with zero-mean and standard deviation $\delta$. Finally, error-convolved magnitudes outside the user-supplied bright and faint limits of the survey are discarded.

Figure \ref{curves}  shows examples of synthetic light curves produced by \ellisa~for eclipsing binaries of the three types: (upper panels) with periods of 0.4745 d (left) and 0.53 d (right), EB (middle panels) with periods of 0.6062 d (left) and  0.77 d (right) and EA (bottom panels) with periods of 3.2293 d (left) and  5.84 d (right), for the VRI filters. For each eclipsing binary type, two examples are shown to illustrate a very dense ($>$90) number of observations (left panels) and very sparse ($\sim$ 35) number of observations (right panels) time sampling characteristic of different areas of the QUEST \citetalias{Mateu} survey. 

The ELLISA code takes approximately 40 minutes to run in an Intel Xeon(R) CPU E5-2620 v3 \@2.9GHz x24, to produce a light curve library for 1,000 eclipsing binaries.

\subsection{Comparison of the ELLISA period distribution with Kepler}

Here we compare the period distribution obtained by ELLISA with that from the Catalogue of Eclipsing Binaries \citep{Keplerebcat} from the Kepler mission. The Kepler catalogue was produced based on the very high number of observations of the Kepler mission, acquired during a long baseline, so it is a very homogeneous and complete survey of eclipsing binaries and is not expected to be significantly affected by observational biases, making a good reference for comparison. 

\begin{figure}
\begin{center}
\includegraphics[width=\columnwidth]{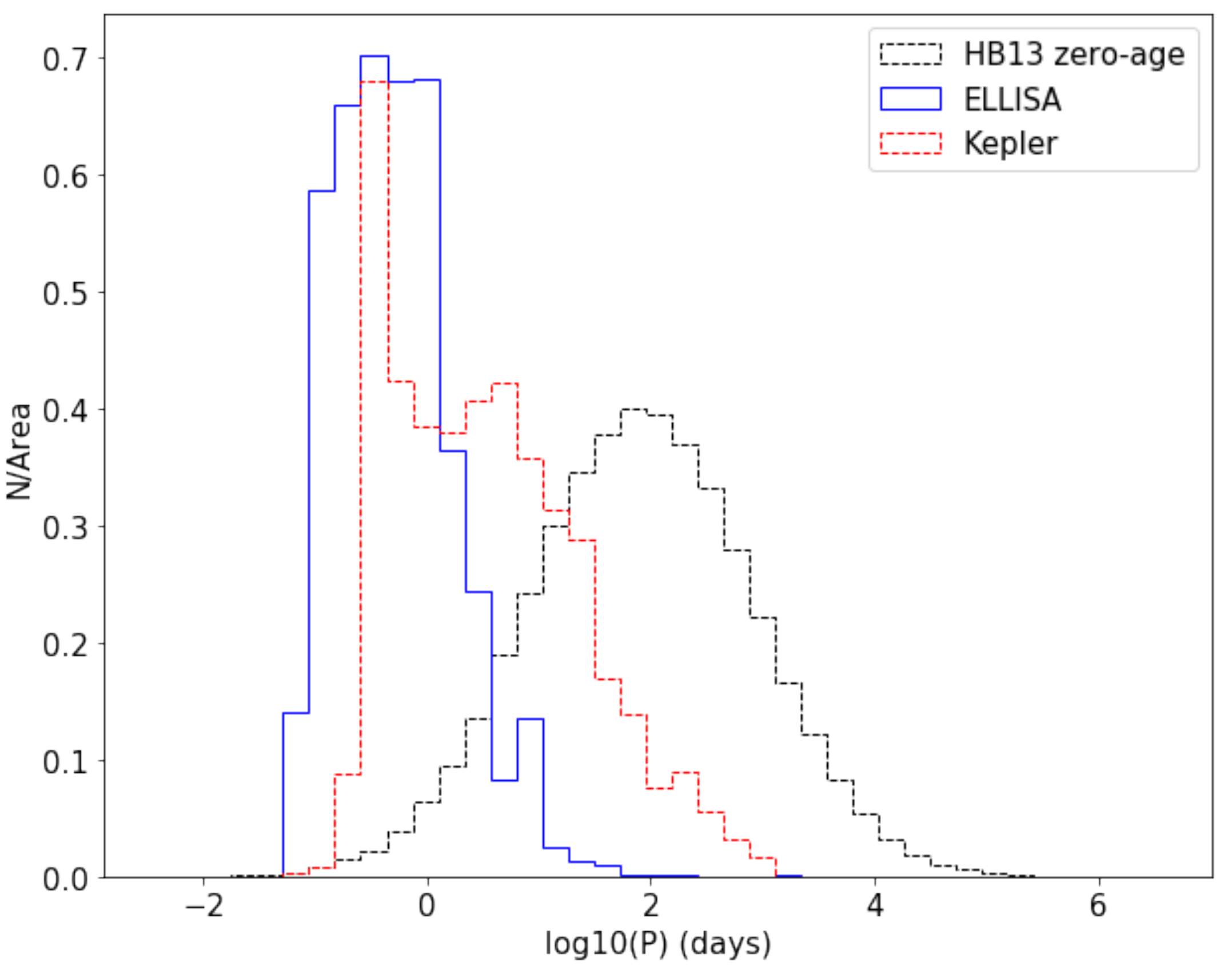}
\caption{Period distribution resulting for a realization of a Thin Disc population of ELLISA eclipsing binaries (solid line), compared to the Kepler eclipsing binaries catalogue \citep{Keplerebcat} (dashed line). The initial period distribution assumed by HB13 (dotted line) for all binaries at zero-age is also shown for reference. All distributions shown are normalised to unit area.}
\label{kepler_ellisa_periods}
\end{center}
\end{figure}

Figure \ref{kepler_ellisa_periods} shows the period distribution 
for an ELLISA simulation of eclipsing binaries in a Thin Disc population, compared to the Kepler distribution. The initial distribution assumed as an ingredient of HB13 in the ELLISA simulation is also shown (dotted) and corresponds to the period distribution of the full binary population at zero-age (see Section~\ref{Sec:parfisicosbassic}). By contrast, the distribution shown for ELLISA corresponds only to the binaries that, at the selected age of the population, turn out to be eclipsing (based on the criterion described in Section \ref{s:phoe_sec}). There are some clear differences between the ELLISA and Kepler period distributions, although the shape is similar overall. At the short-period end, there is a sharp drop off at $\sim0.1$~d while in the Kepler distribution, which happens at a slightly lower period in ELLISA $\sim0.02-0.03$~d. At the long-period end, the drop-off is much shallower in Kepler and while the ELLISA distribution falls-off rapidly at periods larger than $\sim 10$~d. This more notable difference is probably due to the criteria used to decide when a binary is eclipsing or not. Some of this differences also might stem from the choice of initial period distribution made in HB13. In the future we plan to explore different initial assumptions for some of the zero-age orbital parameters and possible introduce distributions that consider joint dependencies \citep[e.g.][]{MoeDS17}. For the time being, since this modifications are beyond the scope of this work, we caution the reader that the use of ELLISA be better restricted to simulating eclipsing binaries with periods $\lesssim 15$~d.

\section{The search for eclipsing binaries in the QUEST low-latitude catalogue of variable stars}\label{binary_search}

In what follows we identify candidate variables in the QUEST low-latitude survey and, over the sample of candidates, we conduct a period search using the available filters simultaneously and identifying possible spurious periods. 

To characterise the completeness of the variable stars identification, we use \ellisa~to generate a mock catalogue to mimic the time sampling and photometric errors of the QUEST survey. We search for candidate variables and determine the fraction of recovered variables using the same technique in terms of apparent magnitude and amplitude.  

\subsection{The QUEST low-latitude survey}\label{s:quest_survey}

In this work we make use of the QUEST variability survey at low galactic latitude from \citetalias{Mateu}.  This catalogue spans a total area of 476 deg$^2$ with 6,513,705 point sources and consists of multi-epoch observations in the VRI filters obtained fully with the J\"urgen Stock Schmidt telescope and the QUEST-I camera \citep{Baltay2002} at the National Astronomical Observatory in Llano del Hato, M\'erida, Venezuela.  The observations correspond to a low Galactic latitude zone in the range $-25^\circ \la b \la 30^\circ$ approximately in direction towards the Galactic anti-centre 190$^ \circ \leqslant l \leqslant 230^ \circ$. 

The time sampling of the survey is quite inhomogeneous as it comprises observations from many different projects with different scientific goals, and spans observations obtained between November~1998 and June~2006. Each star was typically observed once in any given night, with some exceptions of areas surveyed more than once (e.g. around McNeil's nebula in Orion). The typical number of epochs per star is $\sim30$ in V,R and $\sim25$ I respectively, but these can range from $\sim10$ up to $\sim120$--$115$ depending on the area of the sky (see Figure~5 in \citetalias{Mateu} and Figure \ref{distespsint} in Section~\ref{s:mockcat}).

The saturation magnitudes $M_{sat}$ are 14.0 mag for V and R and 13.5 mag for I, the limiting magnitudes $M_{lim}$ are 19.7 mag for V and R and 18.8 mag for I and the completeness magnitudes $M_{com}$ are 18.5 mag for filters V and R and 18.0 mag for filter I. For full details of the survey we refer the reader to \citetalias{Mateu}.

\subsection{The \ellisa-QUEST Mock Light Curve Catalogue}\label{s:mockcat}

Here we describe the catalogue of synthetic light curves produced with \ellisa~to mimic QUEST, which is later used to fine-tune the variable star selection thresholds and period search parameters used in the search for eclipsing binaries. 

First we randomly select $>300,000$ of the stars (5\%) out of the \citetalias{Mateu} QUEST catalogue, in order to use the time sampling of each one as an input for \ellisa. This way we can have a simulation with a time sampling in the VRI bands that is representative of the whole survey and that reproduces the different cadence and number of observations across the survey area. For the QUEST \citetalias{Mateu} survey, this a particularly important point since the time sampling and number of observations per filter is very inhomogeneous across the survey. 

\begin{figure*}
\begin{center}
\includegraphics[width=2.1\columnwidth] {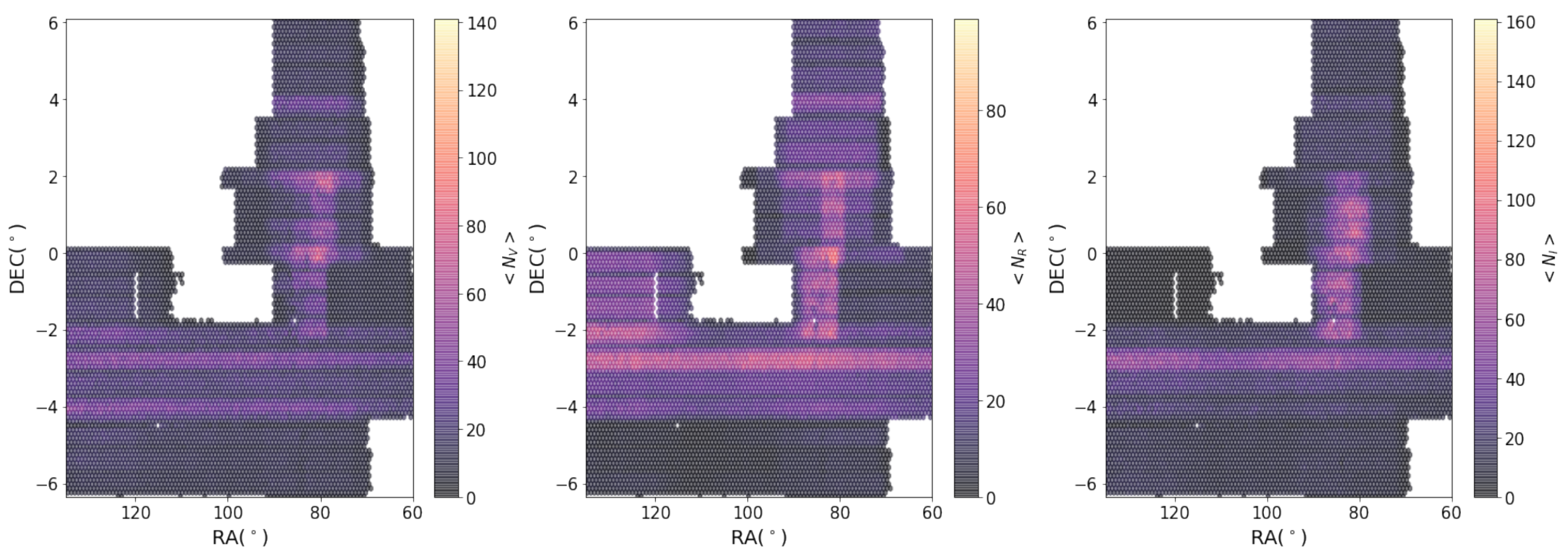}
\caption{Spatial distribution of the number of observations for filters V (left), R (middle), I (right) of the mock catalogue, in equatorial coordinates.  The colour scale in each panel indicates the mean number of observations for each filter for each star.} 
\label{distespsint}
\end{center}
\end{figure*}

The mock light curve catalogue was simulated within the saturation and completeness limits quoted in the previous section for the QUEST survey. Error versus magnitude curves were used as a function of the location in the survey footprint, computed by \citetalias{Mateu} subdividing the whole catalogue in stripes of 1h in right ascension by $0\fdg55$ in declination, allowing to take into account the effect of the changing number of observations per object per filter on the photometric error in different survey regions. Figure~\ref{distespsint} illustrates the number of observations per filter in a map in equatorial coordinates.

The \ellisa-QUEST Mock Light Curve Catalogue was produced using the simulator described in Section~\ref{Sec:ellisa},
assuming a Thin-Disc-like population. This is a reasonable assumption as the \citetalias{Mateu} QUEST survey is concentrated at low galactic latitudes ($|b|<30^\circ$) where the population is expected to be dominated by the Galactic Thin Disc. In any case, at such old ages ($\gtrsim$8--9 Gyr), the overall binary properties change very little with age. So, including a Thick-Disc-like population would be equivalent to giving a slightly larger weight to the old Thin Disc population and should not produce very significant changes.

\begin{figure}
\begin{center}
\includegraphics[width=\columnwidth] {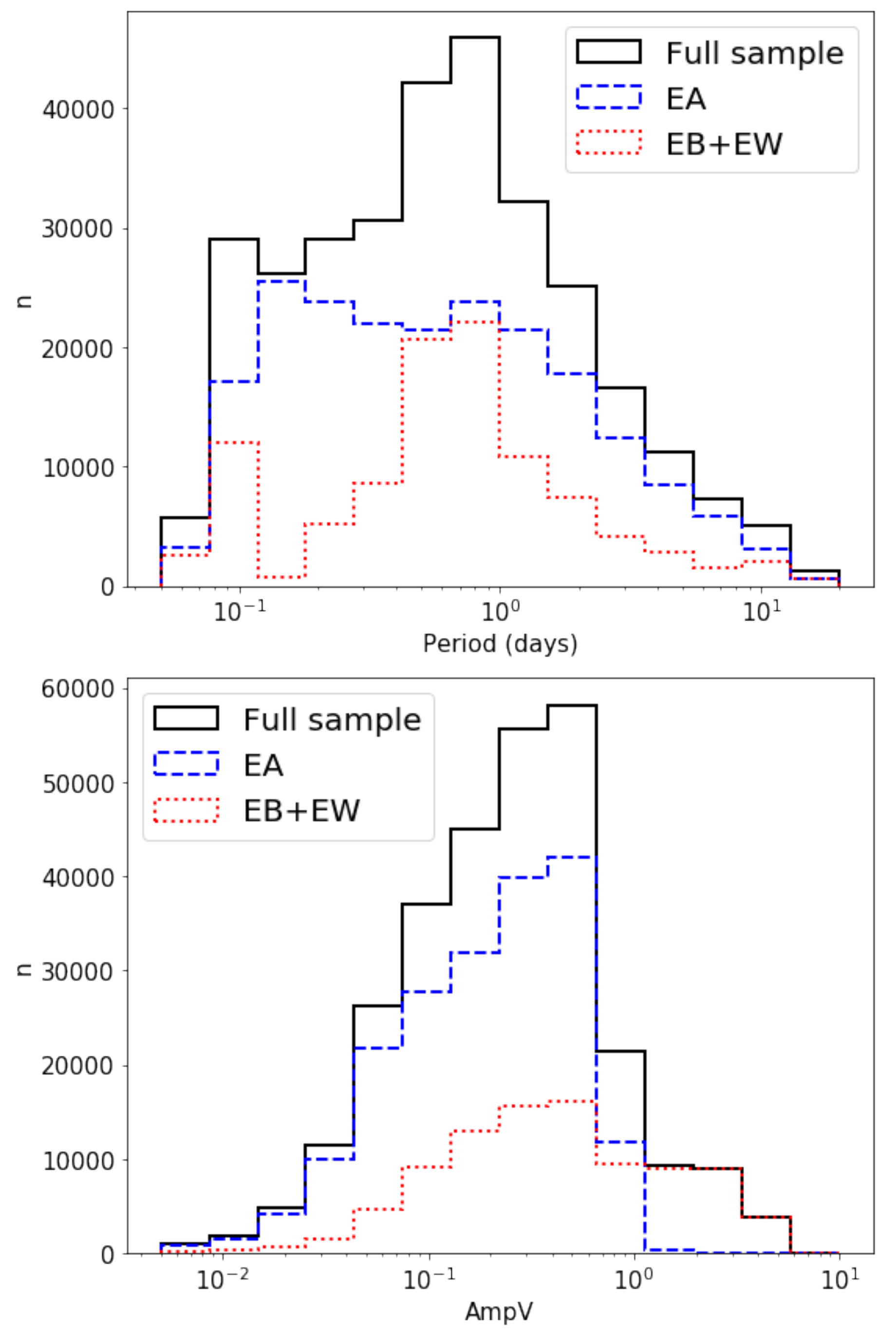}
\caption{Period (upper panel) and V amplitude (lower panel) distribution of full sample (solid black line), EA systems (dashed blue line) and EB+EW systems (dotted red line) of the mock catalogue.} 
\label{dist_per_amp}
\end{center}
\end{figure}

The synthetic light curve catalogue produced containing a total of 307,935 binaries, of which 206,313 are EA systems and 101,622 EB+EW systems.  The period and V band amplitude distributions are shown in the upper and lower panels of Figure \ref{dist_per_amp}, respectively.   
As expected, EA systems dominate the distribution at large periods.  It is interesting to note that, in the amplitude distribution, EB+EW systems dominate at large values ($>1$~mag). This is due to the presence of EW systems composed of massive Main Sequence star pairs, in which stars have large radii ($>\mathrm{few}$~$R_\odot$), common in young populations present in our Disc-like mock population. These binaries are largely absent in old Thick-Disc-like or Halo-like populations (except, e.g., for cases that produce Blue Straggler stars).

The mock catalogue with VRI synthetic light curves is publicly available at the GitHub repository\footnote{\url{https://github.com/umbramortem/QUEST_EBs}}.

In what follows we will use this synthetic or mock light curve catalogue to guide parameters used in the search for eclipsing binaries and to characterise the completeness of the samples at various stages of the identification process. 

\subsection{Variable star identification}\label{s:variables}

The first step in the search for eclipsing binaries is to identify stars exhibiting photometric variability.  To identify variable stars we used the \citetalias{Mateu} extension of the variability indices defined by \cite{Stetson}.  These indices use the fact that the variability in two different photometric passbands must be positively correlated for most known types of variable stars. In the \citetalias{Mateu} catalogue, the three possible Stetson $L$ indices $L_{\rm VR}$, $L_{\rm VI}$ and $L_{\rm RI}$ are reported for each star.

Due to the inhomogeneous coverage of the survey, there are areas with very few observations in one or more filters, or even missing information in one filter altogether. Therefore, as in \citetalias{Mateu}, we selected as variable stars those that meet the following criteria:  

\begin{itemize}
\item{$L_{\rm VR} \geq 1$ or $L_{\rm VI} \geq 1$ or $L_{\rm RI} \geq 1$}
\item{Observed amplitude $\geq 0.2$ mag, at least for two filters}
\item{$N_{obs} \geq 10$, at least for two filters}
\end{itemize}

From the whole QUEST sample of 6,513,705 stars, using this criteria a total of 410,715 stars were selected as candidate variables. We also applied this methodology to the ELLISA-QUEST mock catalogue and 60,145 objects were selected as variable stars.   

\subsection{The period search}

The search for eclipsing variables was conducted using the \cite{Lafler} method developed to identify light curve periods of variable stars. In this work we used the extension presented in \citetalias{Mateu} that allows the use of the multi-filter data in the survey. This method chooses as the best period the one that makes the light curve look smooth when period-folded.  For the purposes of this work, we proceed following \citetalias{Mateu} but adapted the range of the periods search to a wider one considering that the stars studied by \citetalias{Mateu} are RR Lyrae which have a shorter period range than eclipsing binaries.  The period range chosen was from 0.04~d to 15~d, searched with an initial resolution of $10^{-4}$ d, refining the best 5 periods with a resolution of $10^{-7} $~d, and ending with the selection of the 3 best periods.

In addition to the period calculation, we computed the parameter $\Lambda$, a statistical significance parameter that 
describes how smooth the light curve is for a given trial period, compared to the median for all trial periods, in standard deviation units ($N\sigma$). \citetalias{Mateu} used a conservative threshold of $\Lambda=4$ to search for RR Lyrae stars, i.e. stars for which a trial period was found to smooth the light curve at a $4\sigma$ statistical significance.  After a visual inspection of a significant fraction of the sample, we observed that those light curves with values of $\Lambda$ close to 4 were extremely noisy.  Because of this, we selected a higher threshold and performed the visual inspection only for those curves with $\Lambda \geqslant 5$, which resulted in a sample consisting of 10,020 initial candidates. 

\subsubsection{Period aliases}\label{s:aliases}

\cite{Vivas2004} and \citetalias{Mateu} have shown that in the QUEST survey the period aliasing is frequent due to periodicities in the time sampling. Hence, we used the mock catalogue of eclipsing binaries to identify the most common period aliases affecting our survey.

According to \cite{Lafler}, the spurious periods $\Pi$ follow the relation

\begin{equation}
\frac{1}{\Pi}=\frac{k}{P}+\frac{1}{p}
\end{equation}

\noindent with $P$ the true period, $k$ a rational number that produces the harmonics of $P$ and $p$ the external period present in the time sampling. \citetalias{Mateu} analysed the most frequent period aliases that affect the recovery of RR Lyrae stars in their survey. They found the most frequent one is the one-day alias and the less frequent aliases were the 1/3-day and 1/4-day aliases for RRab stars and for the RRc type the first harmonic of the true period and the 1/2-day alias. In this work we have repeated that analysis and summarise our results in Table~\ref{tablaalias}. In addition to the common period aliases found in \citetalias{Mateu}, observe the presence of half the period harmonic aliases with $k=\frac{1}{2}$, as well as another very important locus represented by the 1/20-day (1.5 h) alias. This period alias is interesting as it is probably due to the repeated observations obtained as part of the CIDA variability survey of Orion around the McNeil nebula \citep{Briceno2004}, which had precisely a 1.5 h cadence.  

\begin{table}
\caption{Frequency of occurrence of the different aliases or spurious periods present in the mock catalogue of eclipsing binaries}
\begin{center}
\begin{tabular}{cccc}
\hline
Type        & $k$   & $p$ (d) & Frequency(\%)\\
\hline
Correctly identified & 1 & & 14.3\\
Harmonics & $\frac{1}{2}$ & & 2.79\\
 & 2 & & 16.4\\
 & $\frac{1}{4}$ & & 0.57\\
  & 4 & & 0.53\\
 & $\frac{1}{5}$ & & 0.4\\
  & 5 & & 0.07\\
   & $\frac{1}{20}$ & & 0.37\\
  & 20 & & 0.001\\
1-day alias &	1	&	$\pm$1	&	0.4\\
 &	2	&	$\pm$1	&	1.4\\
$\frac{1}{2}$-day alias & 1 & $\pm\frac{1}{2}$	&	0.1\\
2-day alias &	1	&	$\pm$2	&	0.4\\
$\frac{1}{3}$-day alias &	1	&	$\pm\frac{1}{3}$	&	0.12\\
3-day alias	&	1	&	$\pm$3	&	0.3\\
$\frac{1}{4}$-day alias &	1	&	$\pm\frac{1}{4}$	&	0.1\\
4-day alias	&	1	&	$\pm$4	&	0.17\\
$\frac{1}{5}$-day alias &	1	&	$\pm\frac{1}{5}$	&	0.03\\
5-day alias &	1	&	$\pm$5	&	0.12\\
$\frac{1}{20}$-day alias	&	1	&	$\pm\frac{1}{20}$	&	0.01\\
20-day alias	&	1	&	$\pm$20	& 	0.07\\
\hline
\label{tablaalias}
\end{tabular}
\end{center}
\end{table}

The most common aliases identified in this analysis were used in the visual inspection of the light curves during the search of eclipsing binaries in the real catalogue. Alias periods are reported wherever disambiguation was not possible.

\subsection{The QUEST catalogue of Eclipsing binaries}

The final identification of eclipsing binaries was made by visual inspection of the full sample of 10,020 initial candidates for which a statistically significant period ($\Lambda\geqslant5$) was found. The light curves were period-folded with the three best periods found, and the five most common period aliases identified in Section~\ref{s:aliases} were also checked for the best period selected.

Finally, we obtained 1,125 eclipsing binaries consisting of 179 EA, 60 EB and 886 EW systems corresponding to 16\%, 5\% and 79\% of the catalogue respectively. 

The main light curve parameters are summarised in Table \ref{tab:lc_parameters} and VRI light curves are shown in  Appendix \ref{a:EB_lightcurves}. The VRI time series data are available at a public GitHub repository\footnote{\url{https://github.com/umbramortem/QUEST_EBs}}. 

\begin{table*}
\caption{Basic information and light curve parameters of the eclipsing binaries identified. (This table is published in its entirety as Supporting Information with the electronic version of the article. A portion is shown here for guidance regarding its form and content).}
\label{tab:lc_parameters}
\begin{footnotesize}
\tabcolsep=0.16cm
\begin{tabular}{cccccccccccc}
\hline
ID & RA & DEC & ($N_V$,$N_R$,$N_I$) & Period & AmpV & AmpR & AmpI & Type & V & R & I\\
	& ($^{\circ}$) & ($^{\circ}$) &	& (d)	& (mag) & (mag) & (mag) &	& (mag) & (mag) & (mag) \\
\hline
40430964 & 73.3967 & 1.21155 & (23,24,26) & 0.36275 & 0.48 & 0.50 & 0.45 & EW & 16.52 $\pm$ 0.02 & 16.04 $\pm$ 0.01 & 15.66 $\pm$ 0.01 \\
40440270 & 73.5137 & 1.10068 & (26,25,26) & 0.62019 & 0.41 & 0.46 & 0.41 & EB & 14.85 $\pm$ 0.01 & 14.69 $\pm$ 0.00 & 14.52 $\pm$ 0.01 \\
40451647 & 73.6484 & 1.19106 & (19,25,25) & 0.26241 & 0.62 & 0.62 & 0.64 & EW & 17.52 $\pm$ 0.04 & 17.12 $\pm$ 0.01 & 16.76 $\pm$ 0.02 \\
40507509 & 74.3014 & 1.23649 & (24,25,27) & 0.43003 & 0.21 & 0.26 & 0.28 & EW & 16.56 $\pm$ 0.02 & 16.24 $\pm$ 0.01 & 15.95 $\pm$ 0.01 \\
\hline
\end{tabular}
\end{footnotesize}
\raggedright The Table columns are: the identification number ID, right ascension RA and declination DEC (J2000), the number of observations in the VRI bands, the light curve period, amplitude of variation in each band, type of eclipsing binary and mean magnitudes in each band with the corresponding photometric error.
\end{table*}

\subsubsection{Cross-match with public catalogues of variable stars}

We cross-matched our 1125 Eclipsing Binaries candidates with the ASAS-3 catalogue \citep{Pojmanski}, the GCVS  \citep{Samus} and the Catalina Real-Time Transient Survey (CRTS) \citep{Drake2014} with a tolerance of 7 arcsec. We found no coincidences with ASAS-3, and  found 384 and 12 matches with CRTS and GCVS respectively.  

Out of the 384 matches with CRTS, shown in Table \ref{tab:table_cross2}, 298 stars are classified as EW, 1 as EB and 25 as EA in both catalogues.  Of the 60 matches left, 31 stars are classified as EW in the QUEST catalogue having different classifications in the CRTS catalogue, being 2 EB, 3 EA, 9 RS CVn, 1 RRab, 14 RRc, 1 RRd and 1 Hump (erratic light curve); 19 stars classified as EB in the QUEST catalogue are reported as EA (1) and EW(18) stars in CRTS; and 10 stars classified as EA in the QUEST catalogue and classified as 1 EA with unknown period, 1 EB and 8 EW stars respectively in the CRTS catalogue.  It should be noticed that 12 of the CRTS periods resulting from the cross-match are found to be inexact based on the light curve inspection as indicated by \cite{Drake2014}. We found 180 stars having the same period we obtained in this work within a tolerance of 5\%. These, as well as stars  identified with a period alias are indicated in Table \ref{tab:table_cross2}. Looking at the most common period aliases for the matched stars, we found 9 stars were recovered at half the period, 3 at twice the period and 1 three times the period.  We also found 20 alias of 1-day, 5 of 1/3-day, 11 of 1/2-day and 4 of 1/4-day.

Table~\ref{tab:table_cross1} summarises the matches with GCVS. Out of the 12 matches, 4 stars don't have a computed period in GCVS and are classified there as one rapid irregular in a nebula (NN Ori), one rapid irregular (V1891 Ori), one red rapid irregular(V2678 Ori) and one UV Ceti (V0678 Ori).  These stars were all classified in our catalogue as being EW.

Out of the 8 matches left, 6 were catalogued in GCVS as being RRc stars of which 1 is suspected to be alias of half the period (V0473 Hya),  2 alias of 1-day (V1867 Ori and V1845 Ori) and 3 alias of 1/2-day (V0497 Hya, V0485 Hya and V0944 Mon).  These stars are kept in our final catalogue classified as EW type, since upon visual inspection the curves seem smooth and the amplitude of the eclipses similar for the different filters, as expected for EW binaries. Finally, the two stars left were classified in GCVS as a Delta Scuti with a suspected alias of 1 of 1/3-day (V2742 Ori), an a EW (V2769 Ori).  These stars are classified in our catalogue as EWs.  We retain their classifications as EW since the amplitudes for the three filters are very similar, although we caution that star 51082306 is near the faint limit of our survey ($V=17.6$ at maximum light) and its light curve is noisy.  

\begin{table*}
\caption{QUEST matches with the GCVS catalogue.}
\label{tab:table_cross1}
\begin{tabular}{ccccccccc}
\hline
$\rm{ID}_{QUEST}$ & $P_{\rm QUEST}$ & AmpI & $\rm Type_{QUEST}$ & $\rm ID_{GCVS}$ & $P_{\rm GCVS}$ & AmpI & $\rm Type_{GCVS}$  & Comments\\
 & (d) & (mag) &  & & (d) & (mag) &  & (periods)\\
\hline
50727050 & 0.303348 & 0.442 & EW & V1867 Ori & 0.217985 & 0.5 & RRC: & 1-day alias\\
51082306 & 0.387926 & 0.424 & EW & V2742 Ori & 0.162376 & 0.30000114 & DSCT: & 1/3-day alias\\
80683482 & 1.291147 & 0.468 & EW & V0497 Hya & 0.391934 & 0.6000004 & RRC & 1/2-day alias\\
51556152 & 0.214918 & 0.614 & EW & V2769 Ori & 0.27872 & $\cdots$ & EW & $\cdots$\\
50130978 & 0.337271 & 0.408 & EW & V1845 Ori & 0.254789 & 0.40000057 & RRC: & 1-day alias\\
80312834 & 0.690166 & 0.406 & EW & V0473 Hya & 0.345091 & 0.5 & RRC & k=2 harmonic\\
80010865 & 0.418573 & 0.25 & EW & V0944 Mon & 0.209273 & 0.29999924 & RRC: & 1/2-day alias\\
80494973 & 0.480576 & 0.408 & EW & V0485 Hya & 0.240283 & 0.3000002 & RRC & 1/2-day alias\\
50917935 & 0.474314 & 0.304 & EW & V1891 Ori & $\cdots$ & 0.42000008 & IS & $\cdots$\\
51000045 & 1.66799 & 0.327 & EW & V2678 Ori & $\cdots$ & 0.10999966 & INSB & $\cdots$\\
50765586 & 4.648128 & 0.365 & EW & V0678 Ori & $\cdots$ & 2.0 & UVN & $\cdots$\\
50969201 & 10.696515 & 0.331 & EW & NN Ori & $\cdots$ & 2.500001 & INS & $\cdots$\\
\hline
\end{tabular}
\end{table*}

Our catalogue also overlaps partially with the \cite{Eyken} eclipsing binaries catalogue from the  Palomar Transient Factory (PTF) Orion Project. Out of their total of 82 binaries, we find 22 matching systems,
with 16 of them classified as Close systems (C) which corresponds to EW+EB systems in our classification and 5 as Detached (D) which corresponds to our EA systems, in both catalogues, and 1 system classified in this work as EW and classified as EA system by \cite{Eyken}.  We found 9 period matches, an alias of 1-day and 1 alias of half the period.\\ 

We also cross-matched our catalogue with the T Tauri star catalogue from \cite{Karim2016}, since it is based on the QUEST low-galactic-latitude catalogue with additional observations made by the YETI project for some objects. We found 18 resulting matches, all classified in our catalogue as EW and as 2 Classical T Tauris (CTTS) and 16 weak-lined T Tauris (WTTS) in \cite{Karim2016}. Data for these stars in \cite{Karim2016} catalogue are shown in Table  \ref{tab:table_cross2}, were the identified period aliases are also reported. These matching objects could be potential T Tauri stars in binary systems. However, a more detailed analysis of the light curves would be warranted to confirm this since most are recovered here at twice the period, meaning that only one of the two variability causes is probably real. It is important to note, however, that the low fraction of matching objects on both catalogues, 8\% of \citeauthor{Karim2016}'s and $<$2\% of ours, supports that a low degree of contamination or confusion is expected in either catalogue.

\begin{table*}
\caption{QUEST matches with the CRTS catalogue. (This table is published in its entirety as Supporting Information with the electronic version of the article. A portion is shown here for guidance regarding its form and content)}
\label{tab:table_cross2}
\begin{tabular}{ccccccccc}
\hline
$\rm{ID}_{QUEST}$ & $P_{\rm QUEST}$ & AmpV & $\rm Type_{QUEST}$ & $\rm ID_{CRTS}$ & $P_{\rm CRTS}$ & AmpV & $\rm Type_{CRTS}$  & Comments\\
 & (d) & (mag) &  & & (d) & (mag) &  & (periods)\\
\hline
40047541 & 0.263274 & 0.307 & EW & J041329.6-031402 & 0.26327 & 0.21 & EW & matches\\
40067341 & 0.310505 & 0.329 & EA & J041810.3-024627 & $\cdots$ & 0.07 & EA$_{UP}$ & $\cdots$\\
40083051 & 0.352228 & 0.328 & EW & J042232.8-015739 & 0.352226 & 0.2 & EW & matches\\
40141914 & 0.409572 & 0.523 & EW & J043216.2-015942 & 0.29033 & 0.55 & EW & 1-day alias\\
40152097 & 0.267945 & 0.666 & EW & J043338.3-015918 & 0.23621 & 0.38 & EW & \\
\hline
\end{tabular}
\end{table*}

\begin{table*}
\caption{QUEST matches with the PTF catalogue \protect\citep{Eyken}}
\label{tab:table_cross4}
\begin{tabular}{ccccccc}
\hline
$\rm{ID}_{QUEST}$ & $P_{\rm QUEST}$ & $\rm Type_{QUEST}$ & $\rm ID_{PTF}$ & $P_{\rm PTF}$ & $\rm Type_{PTF}$ & Comments\\
 & (d) & & & (d) & (periods)\\
\hline
50557720 & 0.313152 & EW & 6-3648 & 0.27066 & C & $\cdots$\\
50566490 & 0.540496 & EW & 6-5196 & 0.425248 & C & $\cdots$\\
50571345 & 0.328092 & EW & 6-4262 & 0.328103 & C & matches\\
50747963 & 0.413978 & EW & 9-3480 & 0.342832 & C & $\cdots$\\
50775468 & 0.368418 & EW & 9-6659 & 0.310999 & C & $\cdots$\\
50812538 & 0.086737 & EA & 10-3009 & 2.234 & D & $\cdots$\\
50847483 & 0.468221 & EW & 10-2406 & 0.468237 & C & matches\\
50876465 & 0.327801 & EW & 11-3313 & 0.281535 & C & $\cdots$\\
50879736 & 0.227755 & EW & 11-3778 & 0.227754 & C & matches\\
50620504 & 0.504927 & EW & 7-750 & 0.402915 & C & $\cdots$\\
50707857 & 0.327595 & EW & 8-1414 & 0.281382 & C & $\cdots$\\
50863250 & 0.372698 & EW & 11-1051 & 0.372703 & C & matches\\
50517261 & 0.852987 & EA & 0-8036 & 0.597359 & D & $\cdots$\\
50544270 & 0.26412 & EW & 0-8177 & 0.304414 & C & $\cdots$\\
50563048 & 0.273228 & EW & 0-5197 & 0.273236 & C & matches\\
50820059 & 0.331189 & EW & 4-8796 & 0.397136 & C & $\cdots$\\
50830997 & 0.258346 & EW & 4-7558 & 0.258346 & C & matches\\
50842277 & 0.264612 & EW & 4-9573 & 0.264623 & C & matches\\
50545498 & 3.138798 & EA & 6-7525 & 1.569462 & D & 1-day alias\\
50569288 & 1.351627 & EA & 6-9087 & 1.351641 & D & matches\\
50599532 & 2.071471 & EW & 7-7604 & 2.071092 & D & matches\\
50602623 & 1.91986 & EA & 1-2659 & 0.960161 & D & k=2 harmonic\\
\hline
\end{tabular}
\end{table*}

\begin{table*}
\caption{QUEST matches with the \protect\cite{Karim2016}  TTS catalogue.}
\label{tab:table_cross3}
\begin{tabular}{ccccccc}
\hline
$\rm ID_{QUEST}$ & $P_{\rm QUEST}$ & $\rm Type_{QUEST}$ & $\rm ID_{KARIM}$ & $P_{\rm KARIM}$ & $\rm Type_{KARIM}$ & Comments (periods)\\
 & (d) & & & (d) & \\
\hline
40538170 & 1.793697 & EW & 259 & 0.9 & WTTS & k=2 harmonic\\
50324694 & 2.318376 & EW & 330 & 0.54 & WTTS & 1/2-day alias\\
50493050 & 2.248062 & EW & 431 & 0.11 & WTTS & k=20 harmonic\\
50607212 & 22.161642 & EW & 607 & 11.19 & WTTS & k=2 harmonic\\
50611629 & 2.480004 & EW & 614 & 1.24 & WTTS & k=2 harmonic\\
50627432 & 14.369345 & EW & 200 & 7.18 & WTTS & k=2 harmonic\\
50643673 & 0.795061 & EW & 694 & 0.4 & WTTS & k=2 harmonic\\
50650756 & 0.97408 & EW & 28 & 0.49 & WTTS & 1-day alias\\
50674665 & 9.984332 & EW & 218 & 5.06 & WTTS & k=2 harmonic\\
50737943 & 3.084968 & EW & 968 & 2.82 & WTTS & \\
50765586 & 4.648128 & EW & 1059 & 2.32 & WTTS & k=2 harmonic\\
50765935 & 1.26327 & EW & 1056 & 0.63 & WTTS & k=2 harmonic\\
50936079 & 1.279304 & EW & 129 & 0.64 & CTTS & k=2 harmonic\\
50969201 & 10.696515 & EW & 1593 & 1.23 & WTTS & 1-day alias\\
51033757 & 1.054704 & EW & 154 & 0.53 & WTTS & 1-day alias\\
51052067 & 14.143456 & EW & 1803 & 7.07 & WTTS & k=2 harmonic\\
51053661 & 9.496182 & EW & 1810 & 4.75 & WTTS & k=2 harmonic\\
51293974 & 7.453762 & EW & 1951 & 3.73 & CTTS & k=2 harmonic\\
\hline
\end{tabular}
\end{table*}

\section{Completeness of the QUEST catalogue of eclipsing binaries}\label {s:compsec}

In order to analyse the completeness of the catalogue of eclipsing binaries produced, we used the procedure described in the previous sections to select variable stars and conduct the period search over the \ellisa-QUEST mock catalogue of eclipsing binary light curves.    
We considered as `recovered' all synthetic variables for which the accumulated phase shift $\alpha$ due to the period misidentification is less than 20\%, $|\alpha|<0.2$, where $\alpha$ is given by

\begin{equation}
\alpha = \frac{(P_{recovered} - P_{true})}{P_{true}} \frac{\Delta T}{P_{true}}
\label{alphashift}
\end{equation}

\noindent where $\Delta T$ is the average duration of the time series, weighted by the number of observations in each filter.

\begin{figure}
\begin{center}
\includegraphics[width=0.95\columnwidth]{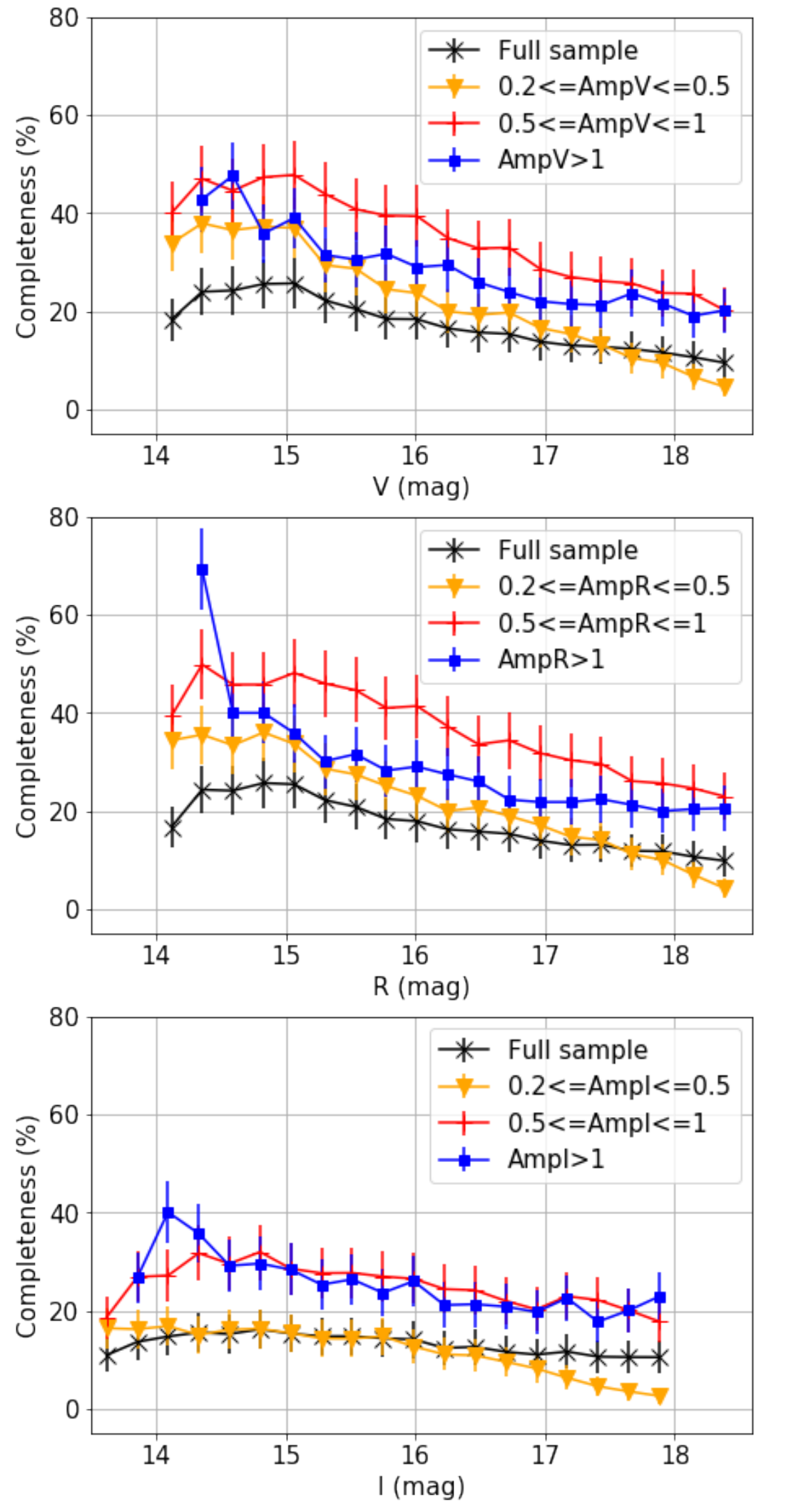}
\caption{Completeness as a function of the V (top), R(centre) and I (bottom) magnitudes of the complete sample (black), stars with amplitudes between 0.2 and 0.5 magnitudes (orange), stars with amplitudes between 0.5 y 1 magnitudes (red) and stars with amplitudes larger than 1 magnitude (blue).}
\label{compvsmag_amp}
\end{center}
\end{figure}

Figure \ref{compvsmag_amp} shows the resulting completeness as a function of the apparent V, R and I band magnitudes. The full sample of synthetic stars with recovered periods corresponds to the black circles. This is sub-divided into three sub-samples according to VRI band amplitudes: $0.2 \leqslant \rm{Amp} \leqslant 0.5$ (yellow triangles), $0.5 \leqslant \rm{Amp} \leqslant 1$ (red diamonds) and $\rm{Amp} \geqslant 1$ (blue squares). Note that, in this and the following figures, the curves represent the completeness or recovery fraction within each sub-sample, so the sum of curves for the different sub-samples \emph{is not equal} to the curve for the combined sample.

For the three ranges selected, the plot shows the completeness decreases monotonically for fainter objects. Also, the recovery is systematically larger for stars in the intermediate-amplitude range, which is to be expected. Lower amplitude stars are more difficult to identify as variables and their periods more difficult to recover, due to the noisier data. Higher amplitude stars, on the other hand, are mostly EA binaries with very abruptly varying light curves for which the \cite{Lafler} method is not optimal, and which are often miss-identified as non-variables, given the difficulty to have well sampled eclipses. Also, note that in the two upper panels the mean completeness for the full sample is lower than for the three sub-samples. This is caused by the stars with data missing in one filter, or an observed amplitude $<0.2$~mag, which are included in the search; this is because in the criteria defined in Section ~\ref{s:variables} we only required two out of the three amplitudes to be larger than 0.2~mag.
 
\begin{figure}
\begin{center}
\includegraphics[width=\columnwidth]{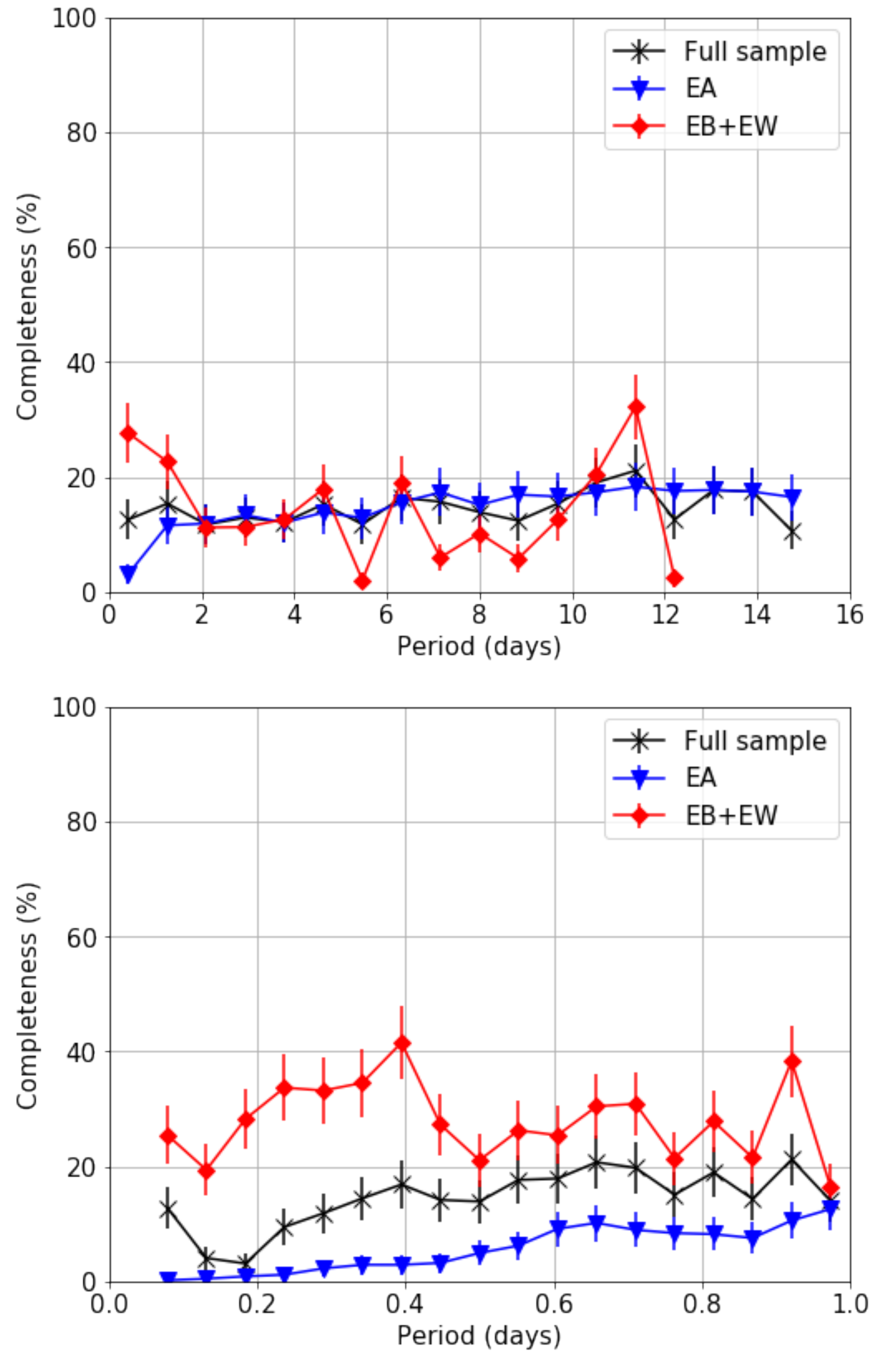}
\caption{Completeness of the identification of eclipsing binaries in two period ranges: for the full period range explored (\emph{upper panel}) and short periods $<1$~d (\emph{bottom panel}). In both panels the average completeness is shown for the full sample (black circles) and by type: EA (blue triangles) and EB+EW (red diamonds).}
\label{comp_period_type}
\end{center}
\end{figure}

Figure~\ref{comp_period_type} shows the completeness as a function of period for each eclipsing binary type, for the full period range (top panel) and zooming in on periods shorter than 1~d (bottom panel). EA binaries are shown with (blue) triangles, EB+EW binaries are shown with (red) diamonds, the combined sample is shown with (black) circles. The top panel shows that taken as a whole, the completeness is on average fairly low, just below $20\%$ for all eclipsing types combined. In this full long-period range going up to $10$~d periods, the EA's completeness is on average $20\%$ which is similar to EB+EW's with an average completeness between $\sim5\%$ and $35\%$. However, this trend is different in the bottom panel focused on periods shorter than 1~d. For this short periods, the EW+EB binaries are optimally recovered with completeness averaging     $\sim 30\%$ for periods larger than $\sim0.3$~d and, in contrast, EA's being recovered less than $5\%$ on average. This confirms our expectation, based on results from the previous RR Lyrae surveys 
\citep[][\citetalias{Mateu}]{Vivas2004}, that EW+EB can be effectively identified with QUEST. The average completeness found for the identification of EB+EW's might seem lower in comparison to the QUEST results for RR Lyrae stars of type $c$, but this is due to the much more stringent recovery criterion used here in comparison to \citet{Vivas2004} and \citetalias{Mateu}. 

\begin{figure}
\begin{center}
\includegraphics[width=\columnwidth]{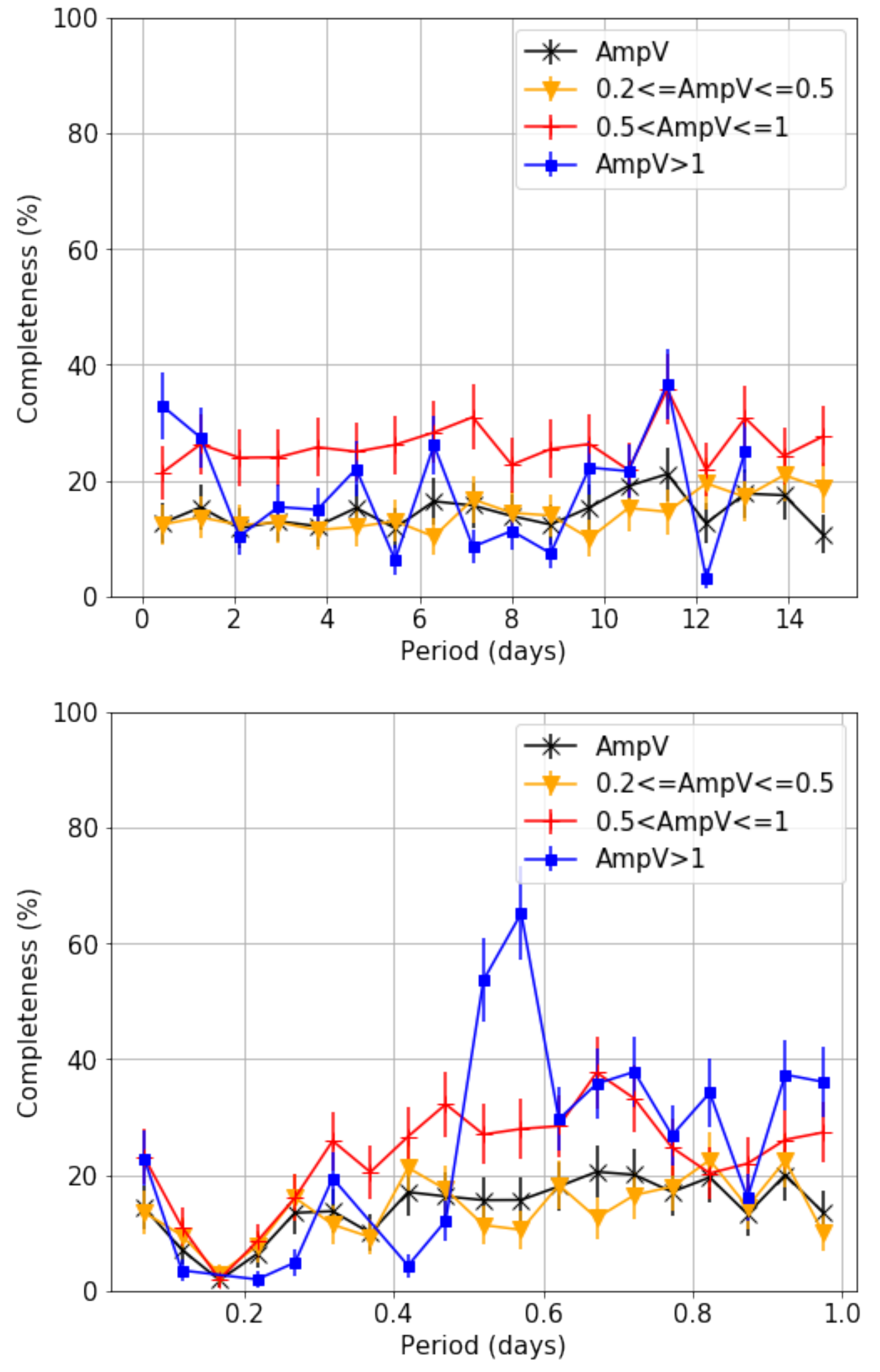}
\caption{Completeness as a function of period for the full period range explored (\emph{upper panel}) and short periods $<1$~d (\emph{bottom panel}), in three ranges of the V-band amplitude. The completeness for the full sample is shown with black circles, and for stars in three amplitude ranges: $0.2 \leqslant \mathrm{AmpV(mag)}\leqslant 0.5$ (yellow triangles), $0.5 \leqslant \mathrm{AmpV(mag)}\leqslant 1.0$ (red diamonds) and $\mathrm{AmpV(mag)}\geqslant 1.0$ (blue squares). The behaviour shown by these plots is also representative of that for the R and I filters (not shown).}
\label{comp_period_ampV}
\end{center}
\end{figure}

Figure~\ref{comp_period_ampV} also illustrates the dependence of the average completeness, independently of the binary type, for the full sample with $\mathrm{AmpV}\geqslant0.2$~mag (black circles) and in three amplitude ranges: short (yellow triangles), intermediate (red diamonds) and large (blue squares). These plots are representative of the behaviour for the R and I amplitudes as well, which are not shown for the sake of simplicity. This figure shows the dependence upon amplitude is not as strong as it is with type, shown in Figure~\ref{comp_period_type}. Binaries in the intermediate amplitude range are slightly better recovered on average, but this is simply a consequence of the correlation of amplitude with period and type, i.e. at a given period the strong dependence shown with type in Figure~\ref{comp_period_type} dominates the average completeness trend more than the amplitude does. This very strong correlation with the eclipsing binary type has to do with the light curve shape and the survey's time sampling: EW+EB's are easier to recover ($<$1~d) because of their smooth continuous light curves, as opposed to EA's sharper eclipses which are harder to sample. Therefore, the main factor that affects completeness is time sampling, rather than amplitude, which plays a secondary role.

\begin{figure*}
\begin{center}
\includegraphics[width=\textwidth]{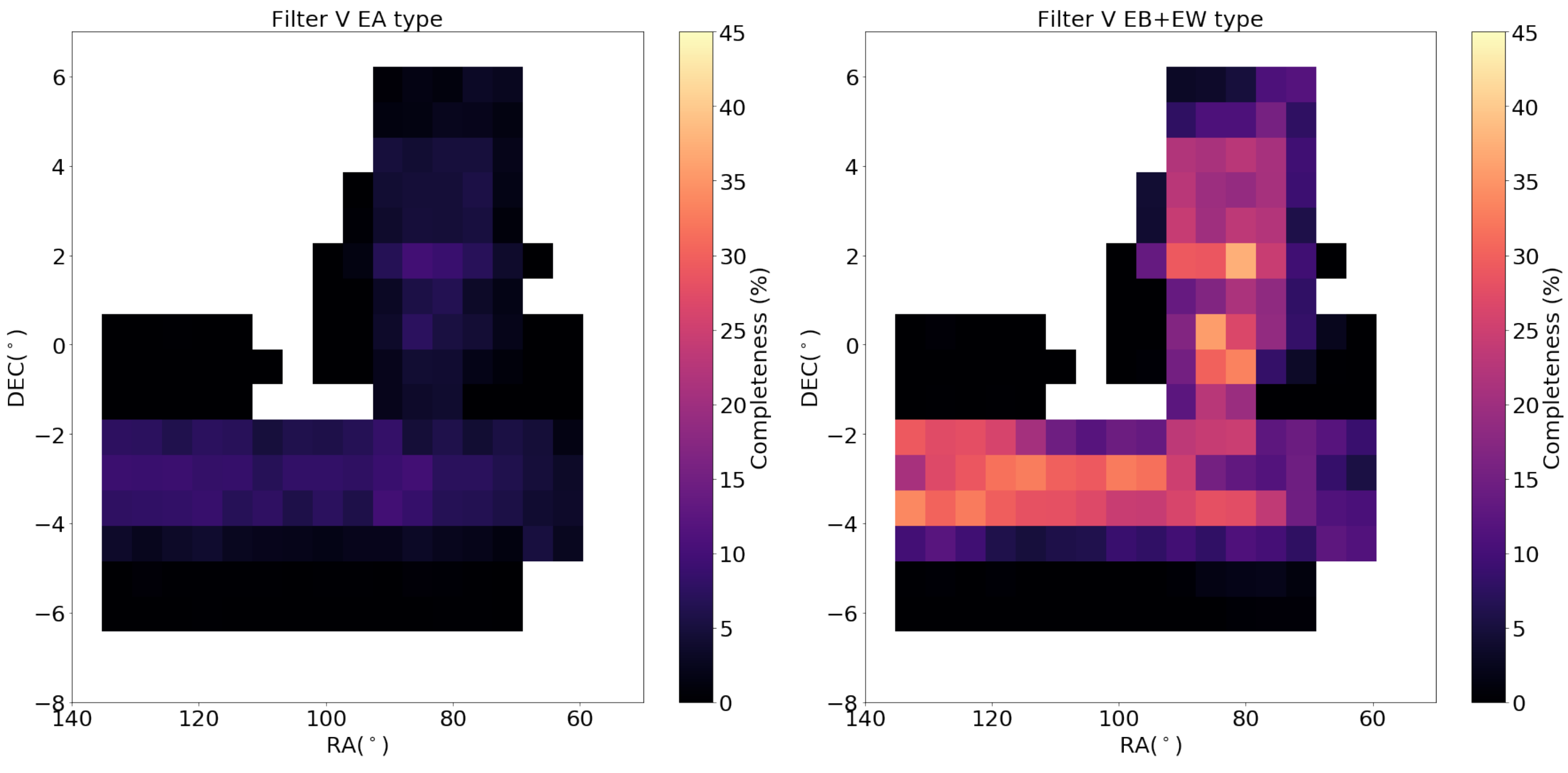}
\caption{Spatial distribution of the average completeness in the recovery of eclipsing binaries (\emph{left:}EA, \emph{right:}EB+EW), in equatorial coordinates. Areas shown in black with null completeness correspond to survey areas with less than 10 epochs per filter on average (see \ref{distespsint}).}
\label{distespsim}
\end{center}
\end{figure*}

Figure \ref{distespsim} shows the completeness as a function of the spatial distribution of  the QUEST low latitude survey. The two panels show the similar overall variations, mainly influenced by the number of available observations as a function of RA-DEC. In the outermost parts of the survey the completeness is lower than that observed for zones with $-5^\circ \leqslant DEC\leqslant +5^\circ$.  The latter, like the zone at $-2^\circ \leqslant DEC \leqslant 0^\circ$ and $60^\circ \leqslant RA \leqslant 80^\circ$, have the lowest number of observations of the survey and also has no observations in R and I (see Figure.  \ref{distespsint}).  On the other hand for the zone between $-4^\circ \leqslant DEC \leqslant -2^\circ$ and $ RA \geqslant 70^\circ$ the completeness is higher than for the previous zones.  We observe that here the number of observations in the different filters is larger than the number for the previous region.  Finally, the highest completeness belongs to the region between 1.5$^\circ \leqslant DEC \leqslant 2^\circ$ and $70^\circ \leqslant RA \leqslant 90^\circ$ which has the larger number of observations for the three filters (typically $>100$ observations per filter).  We conclude that the completeness varies as a function of the number of observations available in each survey region, just as expected, since the fewer observations we have the worst sampled the curves will be, making the period recovery more difficult.  

\section{Summary and Conclusions}\label{sum_and_con}

In this work we introduced \ellisa, a simulator that allows to generate a synthetic library of multi-filter light curves for a population of eclipsing binaries, from user-supplied time sampling and photometric errors.  Mock eclipsing binary light curve catalogues  produced with \ellisa~can be used to test and optimise eclipsing binary searches in upcoming multi-epoch surveys, e.g. \emph{Gaia}, PanSTARRS-1 or LSST, as well as to design observing strategies in future surveys, with a realistic light curve ensemble.
They can also be used to simulate the contamination due to eclipsing binaries in searches for other types of variables, such as RR Lyrae or SX Phoenicis/$\delta$~Scuti stars. \ellisa~is implemented in \textsc{Python} and is publicly available in a GitHub repository\footnote{\url{https://github.com/umbramortem/ELLISA}}.

We have conducted an eclipsing binary search on the QUEST low latitude survey, and used \ellisa~to guide and optimise search parameters for this particular survey, and to estimate the completeness of the resulting eclipsing binary catalogue as a function of binary type, apparent magnitude, period and amplitude. Our main results and conclusions are the following:

\begin{itemize}

\item We have identified 1,125 eclipsing binaries, out of which 707 are new discoveries, and consist of 139 EA (detached or Algol-type), 40 EB (semi-detached or $\beta$Lyrae-type) and 528 EW (contact or  WUMa-type) corresponding respectively to 20\%, 5\% and 75\% of the catalogue.

\item The coordinates and  light curve parameters of the eclipsing binaries identified are summarised in Table~\ref{tab:lc_parameters}. The time series VRI data for the QUEST low latitude eclipsing binary catalogue is made publicly available at a GitHub repository\footnote{\url{https://github.com/umbramortem/QUEST_EBs}}. 

\item EB+EW binaries are identified in QUEST with an average $\sim30\%$ completeness in the period range 0.25--1~d 

\item EAs are identified at an average $15\%$ completeness in the period range $2$--10~d. 

\item Time sampling density is the primary factor affecting which types of eclipsing binary can be recovered and with what average completeness, while amplitude plays a secondary role.

\item This constitutes one of the few catalogues of eclipsing binaries reported to date with a complete characterisation of its selection function, provided as the completeness (recovery fraction) as a function of period, amplitude and apparent magnitude.

\end{itemize}

Being standard candles, EWs can be used to trace Galactic substructure and identify possible new over-densities, streams or dwarf satellites, or to infer Halo and Disc's density profiles, provided the selection function is known, as has been done in numerous previous studies with other standard candles such as RR Lyrae or Red Clump stars \citep{Bovy2016, Mateu2018b,Miceli2008}. The EW catalogue presented here, particularly the high-completeness short-period sample (0.25--1~d), can therefore be used in any of these applications, given the provided characterisation of the catalogue's selection function obtained with the ELLISA mock catalogues.   

\paragraph*{Acknowledgements} The authors are glad to thank the referee, Andrej Pr\v{s}a, his valuable comments and recommendations which helped improve this work. B-CO would like to thank Kyle Conroy for his support with technical details in the implementation of PHOEBE.

\bibliographystyle{mnras}

\bibliography{bibliografia}

\begin{thebibliography}{}
\makeatletter
\relax
\def\mn@urlcharsother{\let\do\@makeother \do\$\do\&\do\#\do\^\do\_\do\%\do\~}
\def\mn@doi{\begingroup\mn@urlcharsother \@ifnextchar [ {\mn@doi@}
  {\mn@doi@[]}}
\def\mn@doi@[#1]#2{\def\@tempa{#1}\ifx\@tempa\@empty \href
  {http://dx.doi.org/#2} {doi:#2}\else \href {http://dx.doi.org/#2} {#1}\fi
  \endgroup}
\def\mn@eprint#1#2{\mn@eprint@#1:#2::\@nil}
\def\mn@eprint@arXiv#1{\href {http://arxiv.org/abs/#1} {{\tt arXiv:#1}}}
\def\mn@eprint@dblp#1{\href {http://dblp.uni-trier.de/rec/bibtex/#1.xml}
  {dblp:#1}}
\def\mn@eprint@#1:#2:#3:#4\@nil{\def\@tempa {#1}\def\@tempb {#2}\def\@tempc
  {#3}\ifx \@tempc \@empty \let \@tempc \@tempb \let \@tempb \@tempa \fi \ifx
  \@tempb \@empty \def\@tempb {arXiv}\fi \@ifundefined
  {mn@eprint@\@tempb}{\@tempb:\@tempc}{\expandafter \expandafter \csname
  mn@eprint@\@tempb\endcsname \expandafter{\@tempc}}}

\bibitem[\protect\citeauthoryear{{Arenou}}{{Arenou}}{2011}]{Arenou}
{Arenou} F.,  2011, in {Docobo} J.~A.,  {Tamazian} V.~S.,   {Balega} Y.~Y.,
  eds,  American Institute of Physics Conference Series Vol. 1346, American
  Institute of Physics Conference Series. pp 107--121,
  \mn@doi{10.1063/1.3597593}

\bibitem[\protect\citeauthoryear{{Baker} \& {Willman}}{{Baker} \&
  {Willman}}{2015}]{Baker2015}
{Baker} M.,  {Willman} B.,  2015, \mn@doi [\aj] {10.1088/0004-6256/150/5/160},
  \href {http://adsabs.harvard.edu/abs/2015AJ....150..160B} {150, 160}

\bibitem[\protect\citeauthoryear{{Baltay} et~al.,}{{Baltay}
  et~al.}{2002}]{Baltay2002}
{Baltay} C.,  et~al., 2002, \mn@doi [\pasp] {10.1086/341705}, \href
  {http://adsabs.harvard.edu/abs/2002PASP..114..780B} {114, 780}

\bibitem[\protect\citeauthoryear{{Bovy}, {Rix}, {Schlafly}, {Nidever},
  {Holtzman}, {Shetrone}  \& {Beers}}{{Bovy} et~al.}{2016}]{Bovy2016}
{Bovy} J.,  {Rix} H.-W.,  {Schlafly} E.~F.,  {Nidever} D.~L.,  {Holtzman}
  J.~A.,  {Shetrone} M.,   {Beers} T.~C.,  2016, \mn@doi [\apj]
  {10.3847/0004-637X/823/1/30}, \href
  {http://adsabs.harvard.edu/abs/2016ApJ...823...30B} {823, 30}

\bibitem[\protect\citeauthoryear{{Brice{\~n}o} et~al.,}{{Brice{\~n}o}
  et~al.}{2004}]{Briceno2004}
{Brice{\~n}o} C.,  et~al., 2004, \mn@doi [\apjl] {10.1086/421395}, \href
  {http://adsabs.harvard.edu/abs/2004ApJ...606L.123B} {606, L123}

\bibitem[\protect\citeauthoryear{{Brown}, {Beers}, {Wilhelm}, {Allende Prieto},
  {Geller}, {Kenyon}  \& {Kurtz}}{{Brown} et~al.}{2008}]{Brown2008}
{Brown} W.~R.,  {Beers} T.~C.,  {Wilhelm} R.,  {Allende Prieto} C.,  {Geller}
  M.~J.,  {Kenyon} S.~J.,   {Kurtz} M.~J.,  2008, \mn@doi [\aj]
  {10.1088/0004-6256/135/2/564}, \href
  {http://adsabs.harvard.edu/abs/2008AJ....135..564B} {135, 564}

\bibitem[\protect\citeauthoryear{{Chabrier}}{{Chabrier}}{2003}]{Chabrier}
{Chabrier} G.,  2003, \mn@doi [\pasp] {10.1086/376392}, \href
  {http://adsabs.harvard.edu/abs/2003PASP..115..763C} {115, 763}

\bibitem[\protect\citeauthoryear{{Cohen}, {Sesar}, {Bahnolzer}, {He},
  {Kulkarni}, {Prince}, {Bellm}  \& {Laher}}{{Cohen} et~al.}{2017}]{Cohen2017}
{Cohen} J.~G.,  {Sesar} B.,  {Bahnolzer} S.,  {He} K.,  {Kulkarni} S.~R.,
  {Prince} T.~A.,  {Bellm} E.,   {Laher} R.~R.,  2017, \mn@doi [\apj]
  {10.3847/1538-4357/aa9120}, \href
  {http://adsabs.harvard.edu/abs/2017ApJ...849..150C} {849, 150}

\bibitem[\protect\citeauthoryear{Drake et~al.,}{Drake et~al.}{2013}]{Drake2013}
Drake A.~J.,  et~al., 2013, \apj, 763, 32

\bibitem[\protect\citeauthoryear{Drake et~al.,}{Drake et~al.}{2014a}]{Drake}
Drake A.,  et~al., 2014a, \apjl, 213, 9

\bibitem[\protect\citeauthoryear{{Drake} et~al.,}{{Drake}
  et~al.}{2014b}]{Drake2014}
{Drake} A.~J.,  et~al., 2014b, \mn@doi [\apjs] {10.1088/0067-0049/213/1/9},
  \href {http://adsabs.harvard.edu/abs/2014ApJS..213....9D} {213, 9}

\bibitem[\protect\citeauthoryear{{Duquennoy} \& {Mayor}}{{Duquennoy} \&
  {Mayor}}{1991}]{Duquennoy}
{Duquennoy} A.,  {Mayor} M.,  1991, \aap, \href
  {http://adsabs.harvard.edu/abs/1991A%26A...248..485D} {248, 485}

\bibitem[\protect\citeauthoryear{{Eggleton}}{{Eggleton}}{2006}]{Eggletonbook}
{Eggleton} P.,  2006, {Evolutionary Processes in Binary and Multiple Stars}

\bibitem[\protect\citeauthoryear{{Etzel}}{{Etzel}}{1981}]{Etzel}
{Etzel} P.~B.,  1981, in {Carling} E.~B.,  {Kopal} Z.,  eds, Photometric and
  Spectroscopic Binary Systems. p.~111

\bibitem[\protect\citeauthoryear{{Gaia Collaboration} et~al.,}{{Gaia
  Collaboration} et~al.}{2017}]{Gaia2017b}
{Gaia Collaboration} et~al., 2017, \mn@doi [\aap]
  {10.1051/0004-6361/201730552}, \href
  {http://adsabs.harvard.edu/abs/2017A%26A...601A..19G} {601, A19}

\bibitem[\protect\citeauthoryear{{Green} et~al.,}{{Green}
  et~al.}{2015}]{Green2015}
{Green} G.~M.,  et~al., 2015, \mn@doi [ApJ] {10.1088/0004-637X/810/1/25}, \href
  {http://adsabs.harvard.edu/abs/2015ApJ...810...25G} {810, 25}

\bibitem[\protect\citeauthoryear{{Han}, {Podsiadlowski}, {Maxted}, {Marsh}  \&
  {Ivanova}}{{Han} et~al.}{2002}]{Han2002}
{Han} Z.,  {Podsiadlowski} P.,  {Maxted} P.~F.~L.,  {Marsh} T.~R.,   {Ivanova}
  N.,  2002, \mn@doi [\mnras] {10.1046/j.1365-8711.2002.05752.x}, \href
  {http://adsabs.harvard.edu/abs/2002MNRAS.336..449H} {336, 449}

\bibitem[\protect\citeauthoryear{{Hern{\'a}ndez-P{\'e}rez} \&
  {Bruzual}}{{Hern{\'a}ndez-P{\'e}rez} \& {Bruzual}}{2013}]{Hernandez}
{Hern{\'a}ndez-P{\'e}rez} F.,  {Bruzual} G.,  2013, \mn@doi [\mnras]
  {10.1093/mnras/stt368}, \href
  {http://adsabs.harvard.edu/abs/2013MNRAS.431.2612H} {431, 2612}

\bibitem[\protect\citeauthoryear{{Hurley}, {Tout}  \& {Pols}}{{Hurley}
  et~al.}{2002}]{Hurley2002}
{Hurley} J.~R.,  {Tout} C.~A.,   {Pols} O.~R.,  2002, \mn@doi [\mnras]
  {10.1046/j.1365-8711.2002.05038.x}, \href
  {http://adsabs.harvard.edu/abs/2002MNRAS.329..897H} {329, 897}

\bibitem[\protect\citeauthoryear{{Ivezi{\'c}}}{{Ivezi{\'c}}}{2010}]{Ivezic2010}
{Ivezi{\'c}} {\v Z}.,  2010, \mn@doi [Highlights of Astronomy]
  {10.1017/S1743921310011956}, \href
  {http://adsabs.harvard.edu/abs/2010HiA....15..817I} {15, 817}

\bibitem[\protect\citeauthoryear{{Kaiser} et~al.,}{{Kaiser}
  et~al.}{2010}]{Kaiser2010}
{Kaiser} N.,  et~al., 2010, in Ground-based and Airborne Telescopes III. p.
  77330E, \mn@doi{10.1117/12.859188}

\bibitem[\protect\citeauthoryear{{Karim} et~al.,}{{Karim}
  et~al.}{2016}]{Karim2016}
{Karim} M.~T.,  et~al., 2016, \mn@doi [\aj] {10.3847/0004-6256/152/6/198},
  \href {http://adsabs.harvard.edu/abs/2016AJ....152..198K} {152, 198}

\bibitem[\protect\citeauthoryear{Keller, Murphy, Prior, Costa  \&
  Schmidt}{Keller et~al.}{2008}]{Keller2008}
Keller S.~C.,  Murphy S.,  Prior S.,  Costa G.~D.,   Schmidt B.,  2008, \mn@doi
  [\apj] {10.1086/526516}, 678, 851

\bibitem[\protect\citeauthoryear{Kinman \& Brown}{Kinman \&
  Brown}{2010}]{Kinman2010}
Kinman T.~D.,  Brown W.~R.,  2010, \mn@doi [The Astronomical Journal]
  {10.1088/0004-6256/139/5/2014}, 139, 2014

\bibitem[\protect\citeauthoryear{{Kirk} et~al.,}{{Kirk}
  et~al.}{2016}]{Keplerebcat}
{Kirk} B.,  et~al., 2016, \mn@doi [\aj] {10.3847/0004-6256/151/3/68}, \href
  {http://adsabs.harvard.edu/abs/2016AJ....151...68K} {151, 68}

\bibitem[\protect\citeauthoryear{{Lada}}{{Lada}}{2006}]{Lada2006}
{Lada} C.~J.,  2006, \mn@doi [\apjl] {10.1086/503158}, \href
  {http://adsabs.harvard.edu/abs/2006ApJ...640L..63L} {640, L63}

\bibitem[\protect\citeauthoryear{{Lafler} \& {Kinman}}{{Lafler} \&
  {Kinman}}{1965}]{Lafler}
{Lafler} J.,  {Kinman} T.~D.,  1965, \mn@doi [\apjs] {10.1086/190116}, \href
  {http://adsabs.harvard.edu/abs/1965ApJS...11..216L} {11, 216}

\bibitem[\protect\citeauthoryear{{Luck}, {Kovtyukh}  \& {Andrievsky}}{{Luck}
  et~al.}{2006}]{Luck2006}
{Luck} R.~E.,  {Kovtyukh} V.~V.,   {Andrievsky} S.~M.,  2006, \mn@doi [\aj]
  {10.1086/505687}, \href {http://adsabs.harvard.edu/abs/2006AJ....132..902L}
  {132, 902}

\bibitem[\protect\citeauthoryear{{Mateu} \& {Vivas}}{{Mateu} \&
  {Vivas}}{2018}]{Mateu2018b}
{Mateu} C.,  {Vivas} A.~K.,  2018, arXiv, \href
  {http://adsabs.harvard.edu/abs/2018arXiv180207798M} {1711.03967}

\bibitem[\protect\citeauthoryear{{Mateu}, {Vivas}, {Downes}, {Brice{\~n}o},
  {Zinn}  \& {Cruz-Diaz}}{{Mateu} et~al.}{2012}]{Mateu}
{Mateu} C.,  {Vivas} A.~K.,  {Downes} J.~J.,  {Brice{\~n}o} C.,  {Zinn} R.,
  {Cruz-Diaz} G.,  2012, \mn@doi [\mnras] {10.1111/j.1365-2966.2012.21968.x},
  \href {http://adsabs.harvard.edu/abs/2012MNRAS.427.3374M} {427, 3374}

\bibitem[\protect\citeauthoryear{Miceli et~al.,}{Miceli
  et~al.}{2008}]{Miceli2008}
Miceli A.,  et~al., 2008, \mn@doi [The Astrophysical Journal] {10.1086/533484},
  678, 865

\bibitem[\protect\citeauthoryear{{Moe} \& {Di Stefano}}{{Moe} \& {Di
  Stefano}}{2017}]{MoeDS17}
{Moe} M.,  {Di Stefano} R.,  2017, \mn@doi [\apjs] {10.3847/1538-4365/aa6fb6},
  \href {http://adsabs.harvard.edu/abs/2017ApJS..230...15M} {230, 15}

\bibitem[\protect\citeauthoryear{{Pojmanski}}{{Pojmanski}}{1997}]{Pojmanski}
{Pojmanski} G.,  1997, \actaa, \href
  {http://adsabs.harvard.edu/abs/1997AcA....47..467P} {47, 467}

\bibitem[\protect\citeauthoryear{{Pr{\v s}a} \& {Zwitter}}{{Pr{\v s}a} \&
  {Zwitter}}{2005}]{Prsa2005}
{Pr{\v s}a} A.,  {Zwitter} T.,  2005, \mn@doi [\apj] {10.1086/430591}, \href
  {http://adsabs.harvard.edu/abs/2005ApJ...628..426P} {628, 426}

\bibitem[\protect\citeauthoryear{{Pr{\v s}a}, {Pepper}  \& {Stassun}}{{Pr{\v
  s}a} et~al.}{2011}]{Prsa2011}
{Pr{\v s}a} A.,  {Pepper} J.,   {Stassun} K.~G.,  2011, \mn@doi [\aj]
  {10.1088/0004-6256/142/2/52}, \href
  {http://adsabs.harvard.edu/abs/2011AJ....142...52P} {142, 52}

\bibitem[\protect\citeauthoryear{{Pr{\v{s}}a} \& {Zwitter}}{{Pr{\v{s}}a} \&
  {Zwitter}}{2005}]{Phoebe}
{Pr{\v{s}}a} A.,  {Zwitter} T.,  2005, \mn@doi [\apj] {10.1086/430591}, \href
  {https://ui.adsabs.harvard.edu/#abs/2005ApJ...628..426P} {628, 426}

\bibitem[\protect\citeauthoryear{{Reggiani} \& {Meyer}}{{Reggiani} \&
  {Meyer}}{2013}]{Reggiani2013}
{Reggiani} M.,  {Meyer} M.~R.,  2013, \mn@doi [\aap]
  {10.1051/0004-6361/201321631}, \href
  {http://adsabs.harvard.edu/abs/2013A%26A...553A.124R} {553, A124}

\bibitem[\protect\citeauthoryear{{Reggiani}, {Robberto}, {Da Rio}, {Meyer},
  {Soderblom}  \& {Ricci}}{{Reggiani} et~al.}{2011}]{Reggiani2011}
{Reggiani} M.,  {Robberto} M.,  {Da Rio} N.,  {Meyer} M.~R.,  {Soderblom}
  D.~R.,   {Ricci} L.,  2011, \mn@doi [\aap] {10.1051/0004-6361/201116946},
  \href {http://adsabs.harvard.edu/abs/2011A%26A...534A..83R} {534, A83}

\bibitem[\protect\citeauthoryear{{Robin} et~al.,}{{Robin} et~al.}{2012}]{Robin}
{Robin} A.~C.,  et~al., 2012, \mn@doi [\aap] {10.1051/0004-6361/201118646},
  \href {http://adsabs.harvard.edu/abs/2012A%26A...543A.100R} {543, A100}

\bibitem[\protect\citeauthoryear{{Rucinski}}{{Rucinski}}{1994}]{Rucinskipcolor}
{Rucinski} S.~M.,  1994, \mn@doi [\pasp] {10.1086/133401}, \href
  {http://adsabs.harvard.edu/abs/1994PASP..106..462R} {106, 462}

\bibitem[\protect\citeauthoryear{{Rucinski}}{{Rucinski}}{1996}]{Rucinski1996}
{Rucinski} S.~M.,  1996, in {Milone} E.~F.,  {Mermilliod} J.-C.,  eds,
  Astronomical Society of the Pacific Conference Series Vol. 90, The Origins,
  Evolution, and Destinies of Binary Stars in Clusters. p.~270 (\mn@eprint {}
  {astro-ph/9508043})

\bibitem[\protect\citeauthoryear{{Rucinski}}{{Rucinski}}{2004}]{Rucinski2004}
{Rucinski} S.~M.,  2004, \mn@doi [\nar] {10.1016/j.newar.2004.03.005}, \href
  {http://adsabs.harvard.edu/abs/2004NewAR..48..703R} {48, 703}

\bibitem[\protect\citeauthoryear{{Rucinski} \& {Duerbeck}}{{Rucinski} \&
  {Duerbeck}}{1997}]{Rucinski1997}
{Rucinski} S.~M.,  {Duerbeck} H.~W.,  1997, in {Bonnet} R.~M.,  et~al., eds,
  ESA Special Publication Vol. 402, Hipparcos - Venice '97. pp 457--460

\bibitem[\protect\citeauthoryear{{Sale} et~al.,}{{Sale}
  et~al.}{2014}]{Sale2014}
{Sale} S.~E.,  et~al., 2014, \mn@doi [MNRAS] {10.1093/mnras/stu1090}, \href
  {http://adsabs.harvard.edu/abs/2014MNRAS.443.2907S} {443, 2907}

\bibitem[\protect\citeauthoryear{{Samus}, {Durlevich}  \& {et al.}}{{Samus}
  et~al.}{2009}]{Samus}
{Samus} N.~N.,  {Durlevich} O.~V.,   {et al.} 2009, VizieR Online Data Catalog,
  \href {http://adsabs.harvard.edu/abs/2009yCat....102025S} {1}

\bibitem[\protect\citeauthoryear{{Sesar} et~al.,}{{Sesar}
  et~al.}{2010a}]{Sesar}
{Sesar} B.,  et~al., 2010a, \mn@doi [\apj] {10.1088/0004-637X/708/1/717}, \href
  {http://adsabs.harvard.edu/abs/2010ApJ...708..717S} {708, 717}

\bibitem[\protect\citeauthoryear{{Sesar}, {Vivas}, {Duffau}  \&
  {Ivezi{\'c}}}{{Sesar} et~al.}{2010b}]{Sesar2010p}
{Sesar} B.,  {Vivas} A.~K.,  {Duffau} S.,   {Ivezi{\'c}} {\v Z}.,  2010b,
  \mn@doi [\apj] {10.1088/0004-637X/717/1/133}, \href
  {http://adsabs.harvard.edu/abs/2010ApJ...717..133S} {717, 133}

\bibitem[\protect\citeauthoryear{{Sesar} et~al.,}{{Sesar}
  et~al.}{2017}]{Sesar2017b}
{Sesar} B.,  et~al., 2017, \mn@doi [\aj] {10.3847/1538-3881/aa661b}, \href
  {http://adsabs.harvard.edu/abs/2017AJ....153..204S} {153, 204}

\bibitem[\protect\citeauthoryear{{Smith} et~al.,}{{Smith}
  et~al.}{2014}]{Smith2014}
{Smith} R.~M.,  et~al., 2014, in Ground-based and Airborne Instrumentation for
  Astronomy V. p. 914779, \mn@doi{10.1117/12.2070014}

\bibitem[\protect\citeauthoryear{{Stetson}}{{Stetson}}{1996}]{Stetson}
{Stetson} P.~B.,  1996, \mn@doi [\pasp] {10.1086/133808}, \href
  {http://adsabs.harvard.edu/abs/1996PASP..108..851S} {108, 851}

\bibitem[\protect\citeauthoryear{{Torrealba} et~al.,}{{Torrealba}
  et~al.}{2015}]{Torrealba2015}
{Torrealba} G.,  et~al., 2015, \mn@doi [\mnras] {10.1093/mnras/stu2274}, \href
  {http://adsabs.harvard.edu/abs/2015MNRAS.446.2251T} {446, 2251}

\bibitem[\protect\citeauthoryear{{Vivas} \& {Zinn}}{{Vivas} \&
  {Zinn}}{2006}]{Vivas2006}
{Vivas} A.~K.,  {Zinn} R.,  2006, \mn@doi [\aj] {10.1086/505200}, \href
  {http://adsabs.harvard.edu/abs/2006AJ....132..714V} {132, 714}

\bibitem[\protect\citeauthoryear{Vivas, Zinn, Abad, Andrews, Bailyn, Baltay,
  Bongiovanni  \& et. al}{Vivas et~al.}{2004}]{Vivas2004}
Vivas A.~K.,  Zinn R.,  Abad C.,  Andrews P.,  Bailyn C.,  Baltay C.,
  Bongiovanni A.,   et. al C.~B.,  2004, \mn@doi [The Astronomical Journal]
  {10.1086/380929}, 127, 1158

\bibitem[\protect\citeauthoryear{{Westera}, {Lejeune}, {Buser}, {Cuisinier}  \&
  {Bruzual}}{{Westera} et~al.}{2002}]{Westera2002}
{Westera} P.,  {Lejeune} T.,  {Buser} R.,  {Cuisinier} F.,   {Bruzual} G.,
  2002, \mn@doi [\aap] {10.1051/0004-6361:20011493}, \href
  {http://adsabs.harvard.edu/abs/2002A%26A...381..524W} {381, 524}

\bibitem[\protect\citeauthoryear{{Wichmann}}{{Wichmann}}{2011}]{Wichmann2011}
{Wichmann} R.,  2011, {Nightfall: Animated Views of Eclipsing Binary Stars},
  Astrophysics Source Code Library (\mn@eprint {ascl} {1106.016})

\bibitem[\protect\citeauthoryear{{Wilson} \& {Devinney}}{{Wilson} \&
  {Devinney}}{1971}]{Wilson}
{Wilson} R.~E.,  {Devinney} E.~J.,  1971, \mn@doi [\apj] {10.1086/150986},
  \href {http://adsabs.harvard.edu/abs/1971ApJ...166..605W} {166, 605}

\bibitem[\protect\citeauthoryear{{van Eyken} et~al.,}{{van Eyken}
  et~al.}{2011}]{Eyken}
{van Eyken} J.~C.,  et~al., 2011, \mn@doi [\aj] {10.1088/0004-6256/142/2/60},
  \href {http://adsabs.harvard.edu/abs/2011AJ....142...60V} {142, 60}

\makeatother
\end{thebibliography}

\appendix

\section{Eclipsing Binary light curves}\label{a:EB_lightcurves}
 
\begin{figure*}
\begin{center}
\includegraphics[width=1.0\textwidth]{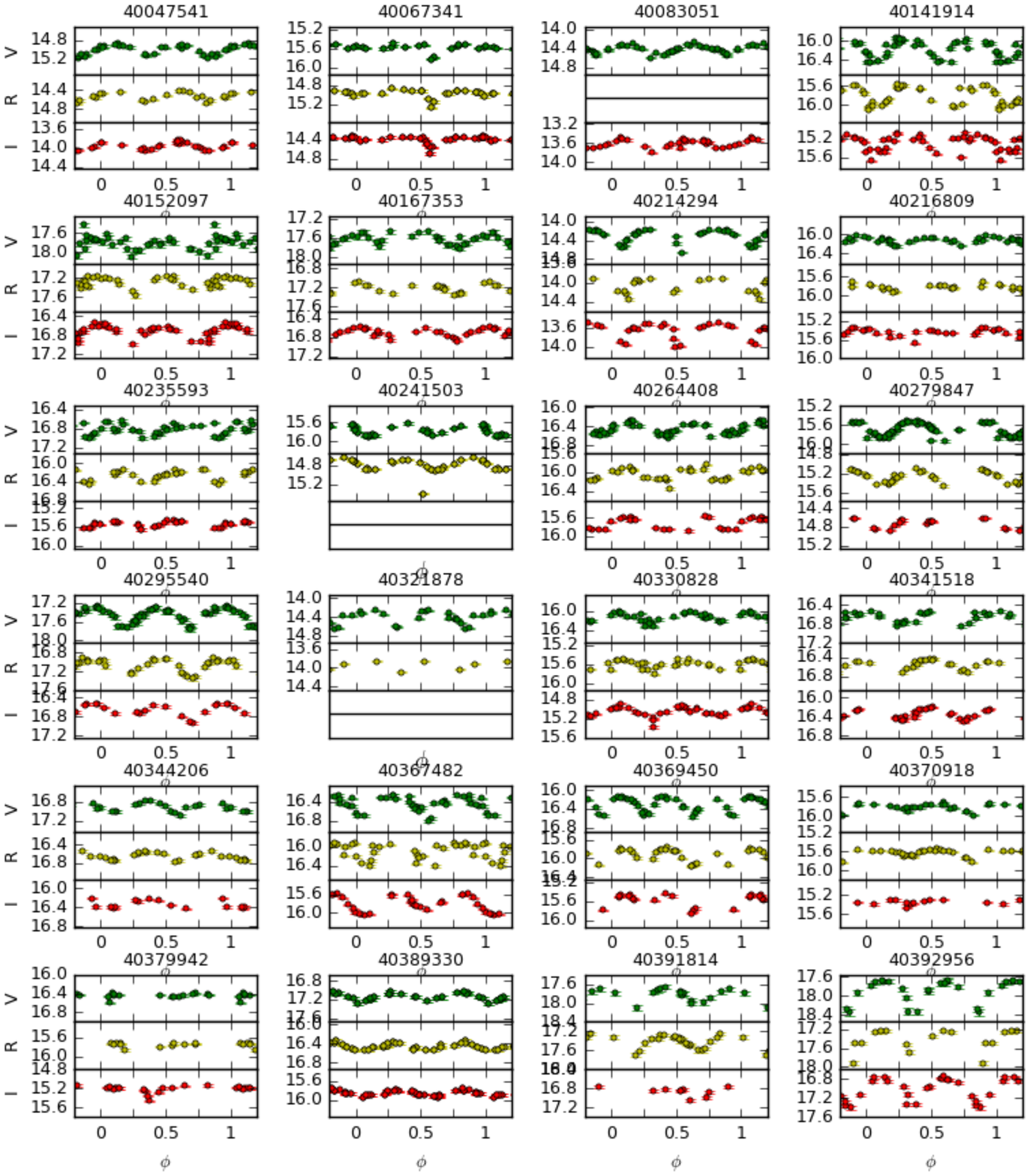} 
\caption{Light curves of survey eclipsing binaries. (This table is published in its entirety as Supporting Information with the electronic version of the article. A portion is shown here for guidance regarding its form and content.)} \label{f:eb_lcs}
\end{center}
\end{figure*}

The catalogue of V, R and I period-folded light curves for all eclipsing binaries found in the survey is shown in  Figure~\ref{f:eb_lcs}. 

\label{lastpage}

\end{document}